\documentclass[a4paper,fleqn,usenatbib]{mnras}
\pdfoutput=1
\usepackage{natbib}
\usepackage{wrapfig}
\usepackage{footnote}
\usepackage{newtxtext,newtxmath}
\usepackage{amsmath}
\usepackage{graphicx}
\usepackage{booktabs}
\usepackage[dvipsnames]{xcolor}
\usepackage{array}
\newcolumntype{?}{!{\vrule width 1.8pt}}
\usepackage{epstopdf}
\usepackage{boldline, multirow}
\usepackage{ctable}
\usepackage{colortbl}
\usepackage{float}
\usepackage{afterpage}
\usepackage[T1]{fontenc}
\usepackage{ae,aecompl}
\usepackage{upquote}
\usepackage{hyperref}
\title[Galaxy Mergers in TNG100-1]{Galaxy interactions in IllustrisTNG-100, I: The power and limitations of visual identification}

\author[K. A. Blumenthal et al.]{Kelly A. Blumenthal$^{1,2}$, Jorge Moreno$^{2,3,4}$,
Joshua E. Barnes$^{1}$, 
Lars Hernquist$^{2}$, 
\newauthor Paul Torrey$^{5,6}$, Zachary Claytor$^{1}$, Vicente Rodriguez-Gomez$^{7}$, 
Federico Marinacci$^{8}$,  
\newauthor and Mark Vogelsberger$^{5}$
\\
$^{1}$Institute for Astronomy, University of Hawaii, 2680 Woodlawn Drive, Honolulu, HI 96822, USA \\
$^{2}$Harvard-Smithsonian Center for Astrophysics, 60 Garden Street, Cambridge, MA 02138, USA\\
$^{3}$Department of Physics and Astronomy, Pomona College, 333 N. College Way, Claremont, CA 91711, USA\\
$^{4}$TAPIR, Mailcode 350-17, California Institute of Technology, Pasadena, CA 91125, USA \\
$^{5}$ MIT Kavli Institute for Astrophysics \& Space Research, Cambridge, MA 02139, USA \\
$^{6}$ Department of Astronomy, University of Florida, 211 Bryant Space Sciences Center, Gainesville, FL, USA \\
$^{7}$Instituto de Radioastronom\'ia y Astrof\'isica, Universidad Nacional Aut\'onoma de M\'exico, Apdo. Postal 72-3, 58089 Morelia, Mexico \\
$^{8}$ Department of Physics \& Astronomy, University of Bologna, via Gobetti 93/2, 40129 Bologna, Italy}

\date{Accepted 2019 December 04. Received 2019 November 05; in original form 2019 May 06}

\pubyear{2019}

\begin{document}
\label{firstpage}
\pagerange{\pageref{firstpage}--\pageref{lastpage}}
\maketitle

\begin{abstract}
We present a sample of 446 galaxy pairs constructed using the cosmological simulation IllustrisTNG-100 at $z=0$, with M$_{\rm FoF, \, dm} = 10^{11}-10^{13.5}$ M$_{\odot}$. We produce ideal mock SDSS $g$-band images of all pairs to test the reliability of visual classification schema employed to produce samples of interacting galaxies. We visually classify each image as interacting or not based on the presence of a close neighbour, the presence of stellar debris fields, disturbed discs, and/or tidal features. By inspecting the trajectories of the pairs, we determine that these indicators correctly identify interacting galaxies $\sim$45\% of the time. We subsequently split the sample into the visually identified interacting pairs (VIP; 38 pairs) and those which are interacting but are not visually identified (nonVIP; 47 pairs). We find that VIP have undergone a close passage nearly twice as recently as the nonVIP, and typically have higher stellar masses. 

Further, the VIP sit in dark matter haloes that are approximately 2.5 times as massive, in environments nearly 2 times as dense, and are almost a factor of 10 more affected by the tidal forces of their surroundings than the nonVIP. These factors conspire to increase the observability of tidal features and disturbed morphologies, making the VIP more likely to be identified. Thus, merger rate calculations which rely on stellar morphologies are likely to be significantly biased toward massive galaxy pairs which have recently undergone a close passage.
\end{abstract}



\begin{figure*}
	\includegraphics[width=0.99\textwidth]{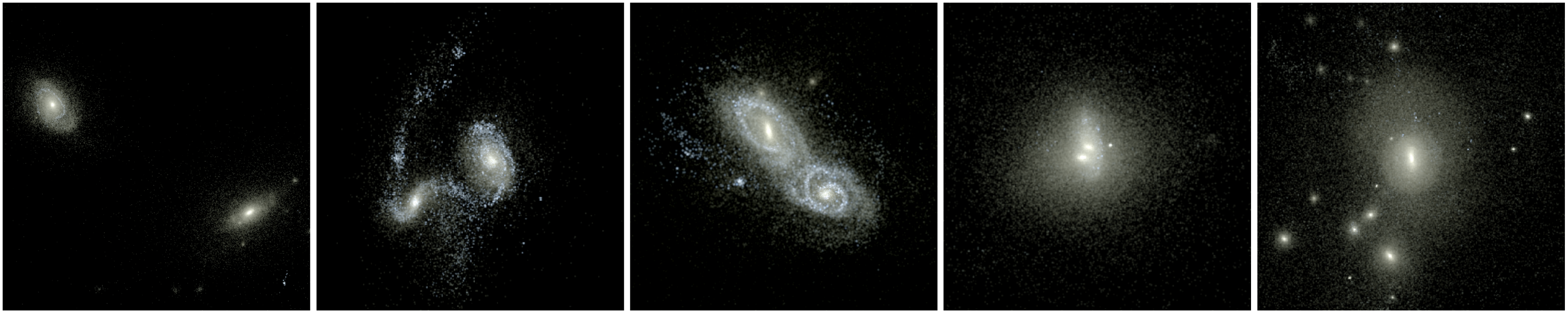}
	\caption{The typical view of the merger sequence includes, from left to right: (1) two galaxies coming in on their initial approach, (2) just after the first pericentric passage, when tidal features are prominent, (3) near second pericentric passage, when there are significant disruptions to the discs and still-visible tidal features, (4)  prior to final coalescence when the galaxies' nuclei are nearly completely overlapping, and (5) a post-merger remnant, featuring clear tidal shells. These simulated three-color composite images are produced via the same procedure described in \S \ref{sec:sdssims} utilising SDSS $g$, $r$ and $i$ magnitudes.}
	\label{fig:mstages}
\end{figure*}

\section{Introduction}
Galaxy encounters have been used to explain the presence of peculiar galaxies \citep[e.g.,][]{arp1966}, and facilitate our understanding of galaxy evolution in a number of ways. Figure \ref{fig:mstages} illustrates the typical merger sequence, from initial approach (far left) to final coalescence (far right). These encounters lead to significant changes in stellar and gas morphology \citep[e.g.,][]{mihos1995a, mihos1995b, malin1997, cote1998, knierman2003, lotz2008, wen2016, tapia2017}, including the production of non-axisymmetric torques which enable gaseous inflows \citep[e.g.,][]{duc2004, blumenthal2018}, which may feed the central black hole, producing heightened activity of the nucleus \citep[e.g.,][]{cutri1985, dahari1985, heck1986a, heck1986b, ellison2011b, hewlett2017, trak2017}. Interacting and merging galaxies have been shown to host heightened rates of star formation \citep[e.g.,][]{joseph1985, kenn1987, whit1995, vig1996, mirabel1998, bridge2007, scudder2012a, moreno2015, rich2015, moreno2019}. In the Local Universe, gas-rich mergers manifest themselves as (ultra)luminous infrared galaxies \citep[e.g.,][]{sanders1988, hopkins2008a}. Until the James Webb Space Telescope begins operations, our knowledge of these objects at high redshifts will remain severely limited. However, indirect measurements such as the observation that early discs are dominated by large clumps of gas and dust \citep[e.g.,][]{lotz2006, ravi2006, whitaker2015} and that many early ellipsoids are very compact \citep[e.g.,][]{buitrago2008, williams2014, barro2013}, have led some to postulate that mergers were much more common in the Early Universe \citep[e.g.,][]{conselice2004, genzel2008, bezanson2009, bour2009, dekel2009}. These findings are consistent with $\Lambda$CDM cosmology, which dictates that the hierarchical structure of the universe arises from sequential mergers throughout cosmic time \citep[e.g.,][]{white1978}.

A fundamental component of galaxy evolution, and by extension hierarchical growth, is the galaxy merger rate. In its simplest form, the galaxy merger rate is calculated by dividing the fraction of galaxies undergoing a merger by the typical time a galaxy interaction will be observable. The merger fraction is often determined by counting the number of morphologically disturbed \citep[both automatically or by visual inspection, e.g.,][]{lotz2008, jogee2009, shi2009, lotz2010, bluck2012}, or the number of galaxies in close pairs \citep[projected or 3D, e.g.,][]{bundy2004, kartaltepe2007, bundy2009, robo2014, mundy2017, snyder2017}. The observability timescale is also variable, and may depend on the orbital parameters and initial conditions \citep[e.g.,][]{conselice2006, lotz2010a, lotz2010}, the observational method used to characterize the merger \citep[e.g.,][]{lotz2008}, and the redshift of the interaction \citep[e.g.,][]{snyder2017}. Due to the breadth of observational methods used to derive these quantities, the calculated merger rate varies widely \citep[e.g.,][]{lotz2011}. However, cosmological simulations are providing insight into the limitations of purely observational studies \citep[e.g.,][]{rod2015}.

In this paper, we set out to answer the following questions that are fundamental to the calculation of the galaxy merger rate across cosmic time:
\begin{enumerate}
	\item Does the stellar morphology of a merging pair reliably indicate its dynamical history?
	\item What makes an interaction ``visible''?
	\item Are merger catalogs derived solely from optical observations biased?
\end{enumerate}

\cite{lotz2011} use small-scale hydrodynamic simulations of binary galaxy pairs to derive a realistic observability timescale. They find that applying this parameter to observational data causes the widely divergent merger rates to converge. \cite{simons2019} use synthetic galaxy images from zoom simulations to determine how frequently galaxies are confused for discs in merger catalogues. These interlopers confuse the disc/spheroid ratio that is often used to define merger rates, a trend the authors found was dependent upon stellar mass. More recently there has been a push to apply deep learning \citep[e.g.,][]{bottrell2019, pearson2019, snyder2019} techniques to synthetic galaxy image catalogues to assess the completeness of observationally derived catalogues. 

In this work we utilise the IllustrisTNG simulation with a volume of $\sim$100$^{3}$ cMpc$^{3}$ (hereafter TNG100-1), one of the three main runs of the IllustrisTNG cosmological suite \citep{marinacci2018b, naiman2018, nelson2018, pill2018b, springel2018}. The IllustrisTNG model \citep{wein2017, pill2018a} employs state-of-the-art prescriptions for star formation, chemical evolution, and feedback due to active galactic nuclei. Recent work has shown that the IllustrisTNG model matches important observational benchmarks in the chemical and metallicty evolution of galaxies \citep[e.g.,][]{naiman2018, torrey2019}, the quasar luminosity function and black hole mass relationships \citep[e.g.,][]{wein2018}, and the overall morphologies of galaxies \citep[e.g.,][]{rod2019}. Using a sample of galaxy pairs from TNG100-1, we generate ideal mock SDSS images to identify what fraction of the interacting pairs are ``observable.''

This paper is structured as follows: in Section \ref{sec:methods} we describe the cosmological simulation used, the methods associated with its data products, and the pair sample preparation; in Section \ref{sec:results} we present and discuss our results, which compare the TNG100-1 sample of interacting pairs at two epochs; finally, in Section \ref{sec:conclusions}, we state our conclusions and briefly describe our future and ongoing work.

\section{Methods}
\label{sec:methods}
\subsection{IllustrisTNG}
IllustrisTNG is a set of N-body/magnetohydrodynamic cosmological simulations with dark and baryonic matter. Gravity is solved using a Tree-PM algorithm that implements a particle mesh on large scales and a tree code on small scales. Gas is treated as an ideal fluid on an unstructured mesh \citep[{\sc AREPO};][]{springel2010a} that incorporates an ideal treatment of magnetohydrodynamics \citep{pakmor2011, pakmor2016}. Gas is allowed to cool via metal-lines and radiation, and can also heat radiatively by exposure to a redshift-dependent radiation field \citep[e.g.,][]{katz1996, faucher2009}. High density gas can self-shield \citep[][]{vogel2013}, under the appropriate optical depth conditions. The interstellar medium (ISM) is modeled with an effective two phase model, following \cite{springel2003b}: cool clouds are in pressure equilibrium with the hot diffuse medium. These simulations cannot describe the ISM structure in detail, but do include mass exchange via cooling, star formation, and the evaporation of clouds by supernovae. This acts to harden the equation of state of the star forming gas, and also stabilizes gas against instability. The ISM prescription does not reach low (high) enough temperatures (densities) to properly describe the molecular gas component. These simulations do not include modeling of cosmic rays nor explicit radiative transfer.

Each star particle represents a stellar population, not an individual star, based on empirical models that include stellar evolution, enrichment, mass and metal returns and supernova rates \citep{pill2018a}. Star formation and supernovae drive outflows in galaxies. Gas mass is ejected from star forming regions such that the wind velocity is proportional to the dark matter velocity dispersion. Due to resolution limitations, outflowing material is initially hydrodynamically decoupled, and is re-coupled at a density threshold. The winds carry a sufficient metal content out of the galaxy, to approximately match the mass-metallicity (or, M-Z) relation \citep[for further details, see][]{pill2018a}.

Black holes (BHs) -- and the feedback due to active galactic nuclei (AGN) -- are a key part of this simulation, in particular the production of quiescent galaxies. Given the resolution of the simulation, black hole formation cannot be self-consistently modeled, so once a galactic halo reaches a certain mass, a seed black hole particle is inserted at its centre, which then acts as a sink particle. The black hole is thus tied to the potential minimum and grows by subsequent mass accretion via Eddington-limited Bondi-Hoyle accretion \citep{springel2005b}. The channel of AGN feedback \citep{sijacki2007, wein2017} depends upon the accretion rate. At low accretion rates, the galaxy experiences a wind (or, kinetic mode), wherein kinetic energy is deposited into the gas around the black hole. The duty cycle then ensures star formation remains suppressed. At high accretion rates, the galaxy enters the thermal (or, quasar) mode; the strength of this feedback mode is a function of the black hole mass. Full details of the IllustrisTNG BH feedback model are available in \cite{wein2017}.

\subsubsection{IllustrisTNG vs. Illustris}
The IllustrisTNG model differs from its earlier counterpart, Illustris \citep{vogel2014a, vogel2014b, genel2014, sijacki2015}, in several ways: (1) it includes isotropic winds with velocities that scale according to the halo virial mass; (2) the supernova energy has two components (thermal and kinetic) which are applied to winds; (3) the wind energy is metallicity-dependent; (4) the supernova mass limit has been set to 8 M$_{\odot}$, and the yield tables have been updated \citep{naiman2018}; (5) it includes an ideal treatment of magnetohydrodynamics. Further, the IllustrisTNG model was run at three different volumes to generate a simulation series that spans a wide dynamical range: TNG50 \citep{nelson2019, pill2019}, TNG100 and TNG300. Each of these runs was initialised with three (TNG100 and TNG300) or four (TNG50) sets of initial conditions, often indicated as e.g., TNG100-1. For more information on the simulation series structure, see Appendix \ref{sec:App1}. In this work, we utilise the run TNG100-1 for several reasons: (1) it has the same set of initial conditions as the original Illustris run; (2) it has the largest number of resolution elements for its volume; and (3) the volume is large enough to contain many examples of interacting galaxies (c.f., TNG50), but not too large that these galaxies are poorly resolved (c.f., TNG300, which typically has a baryon mass resolution on the order of $10^{7}$M$_{\odot}$).

Many parameters and model choices of the IllustrisTNG model were calibrated using observational scaling relations and galaxy properties \citep{pill2018a}. Several works outline the successes of this model. \cite{nelson2018} shows that the colour bimodality, which was absent in the original Illustris, possibly due to the previous implementation of black hole feedback, was present in both TNG100 and TNG300. \cite{rod2019} compares synthetic images from TNG100 to an analogous sample from Pan-STARRS. They find TNG100 to be a significant improvement over the original Illustris suite, particularly with respect to the galaxy morphologies. Additionally, chemical evolution \citep{naiman2018}, galaxy mass-metallicity relations \citep{torrey2019}, and the present day quasar luminosity function \citep{wein2018} are broadly consistent with observations. Despite its relative success, there are still areas of contention between the IllustrisTNG model and the observed universe. For example, TNG100 may underproduce bulge-dominated galaxies, and may overproduce red discs and blue spheroids \citep[e.g.,][]{huertasco2019, rod2019}. The high-redshift quasar luminosity function, driven by the feedback mechanisms employed by supermassive black holes, may be in tension with observations \citep{wein2018, hab2019}. Additionally, it has been suggested that there is contention between the observed and simulation H$_{2}$ content in high redshift galaxies \citep{popping2019}. 

\subsubsection{Friends-of-Friends Groups and Subhaloes}
The Friends-of-Friends (hereafter FoF) algorithm utilizes percolation to construct associated groups of particles \citep{davis1985}. Dark matter particles (or chains of particles) are said to be linked if they are closer than $b\bar{l}$, where $\bar{l}$ is the mean interparticle distance and is related to the simulation's mean number density, $\bar{l} = \bar{n}^{\, -1/3}$. The free parameter $b$ is the ratio between the maximum linking distance and the interparticle separation for a homogeneous system; in IllustrisTNG $b = 0.2$. Taken together, this all represents an approximate density threshold below which particles are not considered associated. The baryonic (gas and stars) material is assigned to a particular FoF group based on the membership of the nearest dark matter particle. Subhaloes, on the other hand, are identified via the \textsc{subfind} algorithm \citep{springel2001}. This iteratively strips away particles that are unbound from the central structure, until a bound system above a certain size remains -- in the case of IllustrisTNG, this is 20 particles. In many cases, as in this work, subhaloes are considered galaxies, while FoF groups may contain pairs or groups of galaxies.

\subsection{Galaxy Pair Samples}
\subsubsection{Parent Sample}
\label{sec:sample}
We select FoF haloes in the most recent snapshot (i.e., $z=0$) with a FoF group total dark matter mass between 10$^{11}$ and 10$^{13.5}$ M$_{\odot}$. Additionally, subhaloes are required to have a total dark matter mass between 10$^{10.5}$ and 10$^{13}$ M$_{\odot}$. These mass cuts ensure that we limit ourselves to well-resolved galaxies and haloes; that we avoid systems in which visual features are driven by environmental, non-merger related processes. Subhaloes with a total dark halo mass less than $10^{10.5}$ M$_{\odot}$ are likely to be poorly resolved in both the dark and baryonic material. To ensure the proper mass resolution of the stars, we place a final limit on the subhalo total stellar mass such that both subhaloes in the pair have a total stellar mass above 10$^{9}$ M$_{\odot}$. We place this restriction on the stellar mass primarily because a preliminary inspection of the images described in Section \ref{sec:sdssims} shows that it is very difficult to identify tidal features in systems with stellar mass below 10$^{9}$ M$_{\odot}$ (see Section \ref{sec:sdssims}.  However, abundance matching \citep[e.g.][]{sawala2015} indicates that this might lead to a sample of galaxies with systematically high stellar masses. No limit is placed on the distance between the subhaloes, although they are required to belong to the same FoF halo. We do not consider pairs that straddle two FoF haloes \citep[as in, e.g.,][]{moreno2012, moreno2013}, and note that these systems are not only relatively rare, but are likely to be unbound (and as such, not orbiting one another). We consider only pairs of galaxies with a stellar mass ratio between unity and 1:4 (``major merger'') at the present day. Lastly, the majority of observations \citep[e.g.,][]{bridge2010, ellison2010, lar2016, ventou2017, mantha2018} and idealised simulations \citep[e.g.,][]{tt1972, barnes1991, barnes1996, dimatteo2008, lotz2008, rupke2010, bour2011, hopkins2013, moreno2015, moreno2019} of galaxy mergers typically assume the system is composed of only two galaxies. In order to approximate this assumption, we required that any tertiary subhalo in the FoF group be at most 1/16 the stellar mass of the primary (or, most massive) halo. It should be noted that this restriction will not exclude all recent minor mergers. There may be systems with strong tidal features at the present day due to low mass ratio interactions in the past. However, observers do not have unlimited knowledge about their targets. By not removing these objects, we remain more closely connected to observational surveys of interacting galaxies. Our final set of galaxies contains 446 binary galaxy pairs at $z=0$.

\subsubsection{Ideal Mock SDSS Images}
\label{sec:sdssims}
We generate ideal mock SDSS images for each of the 446 galaxy pairs in our sample. TNG100 provides magnitudes in eight bands for each star particle, which are calculated using \cite{bc03} (assuming no dust). These include SDSS $g, r, i, z$, Buser $U, B, V$, and Palomar $K$. Here, we generate ideal mock SDSS $g$-band images using all star particles bound to a FoF group. This band was chosen to facilitate future comparisons with wide-field observational surveys. To calculate the luminosity of each star particle, we determine the true SDSS $g$  magnitude
\begin{equation}
M_{\rm true} = M_{\rm obs} - \chi k_{\rm filter}
\end{equation}
where $M_{\rm obs}$ is the TNG100 SDSS $g$-band absolute magnitude, $k_{\rm filter}$ is the filter-dependent first order extinction correction, and $\chi$ is the airmass, assumed to be 1.3 for all SDSS bands in TNG100. The apparent magnitudes are needed to derive the flux:
\begin{equation}
m_{\rm true} = M_{\rm true} + \mu
\end{equation}
The distance modulus, $\mu$, is calculated for each set of galaxy pairs using a set distance of 35 Mpc for every system. 
The flux is then
\begin{equation}
f = 10^{0.4 ( m_{\rm true} - m_{\rm zp} )} 
\end{equation}
where $m_{\rm zp}$ is the zero-point of the desired filter. For our images, we use $k_{\rm filter} = 0.15$, $\chi = 1.3$, and $m_{\rm zp} = 25.11$ \citep{stoughton2002}. We project the three-dimensional distribution of particles onto a flat two-dimensional plane, and apply a 2D Gaussian smoothing function with FWHM equal to the radius of a sphere enclosing the 32 nearest star particles, following \cite{torrey2015}. For simplicity, we use the $x$ and $y$ coordinates to define this plane, and do not assume a location or viewing angle for an observer. Thus the sample represents a random set of orientations with no preferred observing direction. Further the images include no treatment of dust attenuation nor a convolution with the SDSS resolution. This affords us optimal conditions to ``observe'' any tidal features in the mocks. Figure \ref{fig:mstages} contains five (rgb) examples of our ideal mock observations. For the full postage stamp collection of the interacting pairs, refer to Appendix \ref{sec:appb}.

\subsubsection{Visual Classification Scheme}
\label{sec:viz}
The merger sequence is defined by the presence (or absence) of tidal features. \cite{lar2016} devised a merger stage classification scheme that includes non-interacting single galaxies (s), minor mergers (m) and major mergers, ranging from before first pericentric passage through final coalescence and post-merger remnant (M$1-$M$5$) of Ultra-Luminous Infrared Galaxies (ULIRGs). The major merger sequence is as follows (Figure \ref{fig:mstages}): \\
\textbf{M1} - Galaxies are well separated and appear to be on their initial approach. \\
\textbf{M2} - Tidal features (bridges and tails) are clearly visible, and likely just after the first close passage. \\
\textbf{M3} - Two individual nuclei are visible in highly disturbed overlapping discs. The tidal tails are still well defined. \\
\textbf{M4} - The two nuclei have now coalesced, but the tidal debris are still visible. \\
\textbf{M5} - A post-merger remnant, with a diffuse outer shell, and little-to-no evidence of tidal features. \\
Using this merger stage classification as a guide, three of the authors independently classified the pairs as either interacting (roughly, stage M2-M5) or not interacting (s-M1). Visual cues including the projected distance, tidal features, and stellar debris were used. Following Galaxy Zoo \citep[][]{galzoo}, we adopt the group consensus (2 out of 3) as the morphological classification for any given pair. 

Identifying merger stage in this way may be subject to certain pitfalls. For example, the production of tidal features is dependent upon encounter geometry, and can, in some retrograde encounters, be completely absent \citep[e.g.,][]{tt1972, dimatteo2007}. Further, \cite{nevin2019} study the effects of initial conditions on various quantitative morphology measures which are often linked to merger activity. They find that the initial conditions, including viewing angle, may affect the derived quantitative morphologies, thus impacting the perceived merger stage. \cite{dubinski1996, mihos1998, springel1999, barnes2016} perform systematic theoretical studies on the tidal response of interacting galaxies, spanning a wide range of galaxy structures. They find that the visibility  of tidal features additionally depends on the galaxy's internal structure (e.g., the dark halo mass and concentration), and that under certain circumstances, galaxy pairs may not show any obvious signs of interaction.

\cite{lotz2008} use quantitative morphological metrics \citep[e.g., Gini and M$_{20}$ of][]{lotz2004} of simulated galaxy mergers to dive deeper into the idea of time-scales for tidal features. They determine when, over the course of an interaction, the tidal response and subsequent morphological disruption are greatest, and find that galaxies tend to exhibit strong tidal features at first pericentre and near final coalescence. In contrast, the morphologies at intermediate passes are largely consistent with a control sample of isolated galaxies.

In addition to the object's look-back time, internal structure, and encounter geometry, the observer viewing angle can also drastically alter the prominence of tidal features. For example, \cite{pop2018} studied shell stellar debris fields (e.g., their Figure 1, and the last panel of our Figure \ref{fig:mstages}) which they found to be present predominantly in merger remnants. The authors show that whether or not a shell is visible in a particular projection depends on the orbital trajectory of the progenitor system.

Though in this work we do not directly account for the viewing angle, we follow \cite{rod2019} and adopt a fixed projection. This is directly analogous to observational studies; one of the main goals of this paper is to test how well observationally-derived and morphology-based classifications identify interacting systems. Projection will be important for the observability of individual systems' tidal features. However in this work, we look at the properties of the interacting pairs as a whole. In doing so, we sample over a random set of viewing angles, thereby minimizing the biases of individual interacting pairs. Clearly, schema which depend entirely on tidal features in the stellar material might be biased in a variety of ways, thereby affecting the kinds of interactions captured.

Several studies have shown that the internal structure of tidal features is highly dependent upon spatial resolution \citep[e.g.,][]{wetz2007}. However, the broader structure of tidal features is less sensitive to this parameter. For example, \cite{moreno2015} and \cite{moreno2019} perform similar simulations of galaxy interactions: improving the spatial resolution by two orders of magnitude does not affect the presence of visual features. Additionally, we can expect a typical TNG100-1 tidal feature to have about 200 star particles, corresponding to less than 1\% relative error in the feature's resolution.

\begin{table}
\centering
\begin{tabular}[width=4in]{c|c|c|c|}
\cline{3-4}
\multicolumn{2}{c|}{\multirow{2}{*}{}}                                                                                                       & \multicolumn{2}{c|}{\begin{tabular}[c]{@{}c@{}}Does the pair appear\\ to be interacting?\end{tabular}} \\ \cline{3-4} 
\multicolumn{2}{l|}{}                                                                                                                        & Yes & \multicolumn{1}{c|}{No}                                      \\ \hline
\multicolumn{1}{|c|}{\multirow{3}{*}{\begin{tabular}[c]{@{}c@{}}Is the pair \\ actually \\ interacting?\end{tabular}}} & Yes & \textbf{\color[rgb]{0.977,0.488,0.445} 38 } & \multicolumn{1}{c|}{\textbf{\color[rgb]{0.4922,0.0039,0.4922} 47}} \\ \cline{2-4} 
\multicolumn{1}{|c|}{}                                                                                                 & \multirow{2}{*}{No} & \multirow{2}{*}{18} & \multicolumn{1}{c|}{\multirow{2}{*}{214}} \\
\multicolumn{1}{|c|}{} & & & \multicolumn{1}{c|}{}                                        \\ \hline \hline
\multicolumn{3}{|l|}{Recently entered the same group} & 97 \\ \hline
\multicolumn{3}{|l|}{Group} & 22 \\ \hline
\multicolumn{3}{|l|}{Not Orbiting} & 10 \\ \hline \hline
\multicolumn{3}{|l|}{Total Pairs} & 446 \\ \hline
\end{tabular}
\caption{A complete description of the classifications for all pairs in the parent sample. In the top section, we report the results of the combined visual (``Does the pair appear to be interacting?'') and trajectory schema (``Is the pair actually interacting?''), including the VIP and nonVIP which are the subsamples used in this paper. Note that the report from the trajectory is taken to be ``truth.'' The middle section reports the number of pairs which were manually eliminated from the sample.}
\label{table:acnt}
\end{table}

\subsubsection{Trajectory Classification}
\label{sec:int}
Using the Sublink merger tree \citep{rod2015}, we extract the 3D orbital motion of the secondary with respect to the primary. A pair is considered ``interacting'' if it has had at least one close passage, the pair is at or nearing an apocentre (i.e., there is an apparent turnover in the relative separation), and there is apparent orbit decay (i.e., sequentially deeper pericentric passages). For interacting pairs which have had only one close passage, we require the pericentric distance be less than $\sim$150 kpc. In considering the full trajectories, we more reliably remove those galaxies which have merely flown past one another. Whilst it could be argued that such systems have \textit{interacted}, they are not \textit{currently interacting}, and are thus not part of our sample.

\begin{figure*}
\centering
\begin{minipage}{\textwidth}
\centering
	\includegraphics[width=0.8\textwidth]{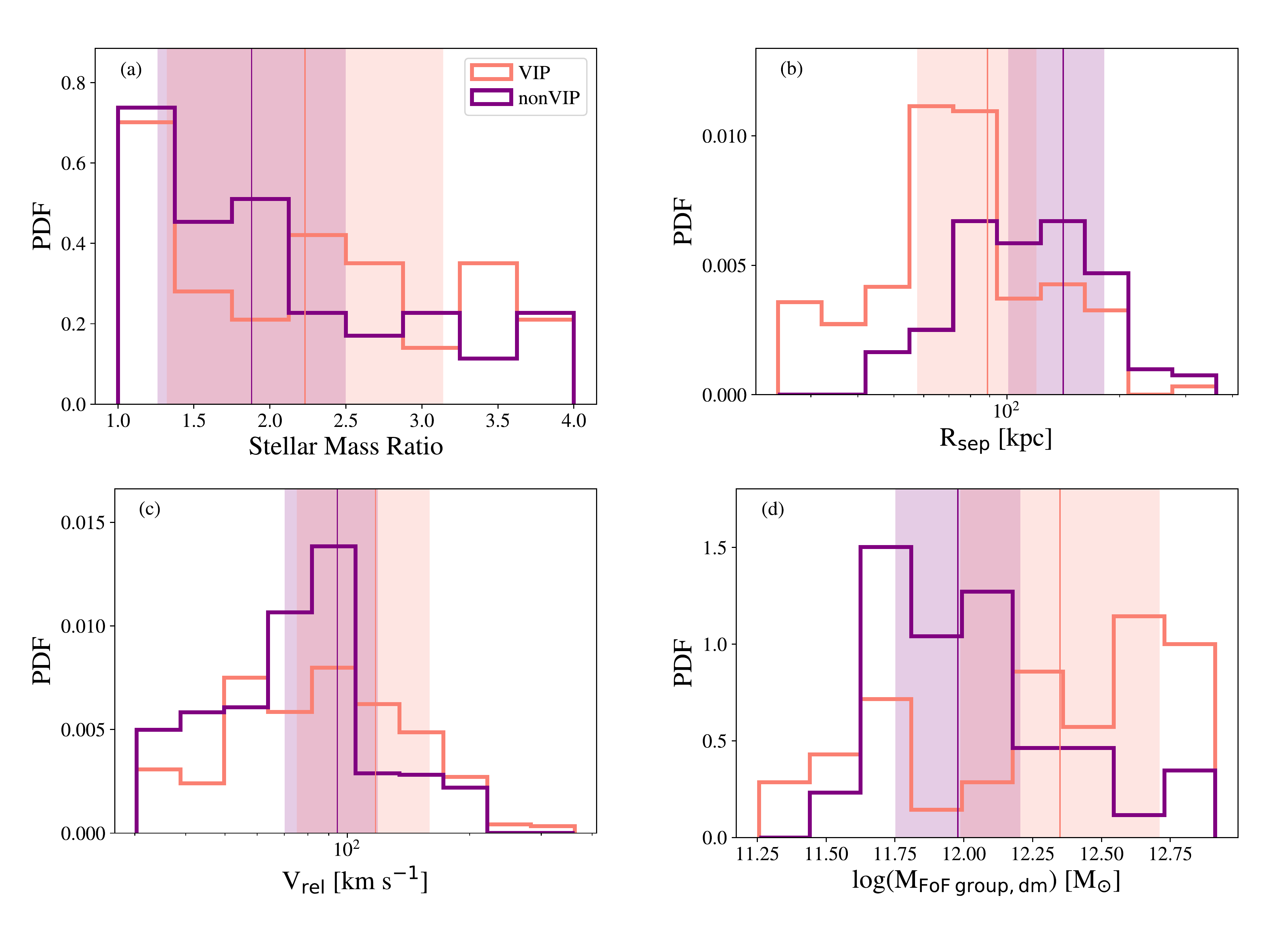}
	\caption{Galaxy-pair property distributions for the Visually Identified Pairs (VIP; salmon) and the non-Visually Identified Pairs (nonVIP; purple). The panels show: (a) the current ($z=0$) stellar mass ratio, (b) $z=0$ 3D separations, (c) relative velocity, and (d) FoF group dark matter mass. The vertical coloured lines correspond to the medians of each sample, and colored rectangles indicate the range within $\pm$ one median absolute deviation from the median.}
	\label{fig:dists_bulk}
\end{minipage}%

\begin{minipage}{\textwidth}
\centering
	\includegraphics[width=0.8\textwidth]{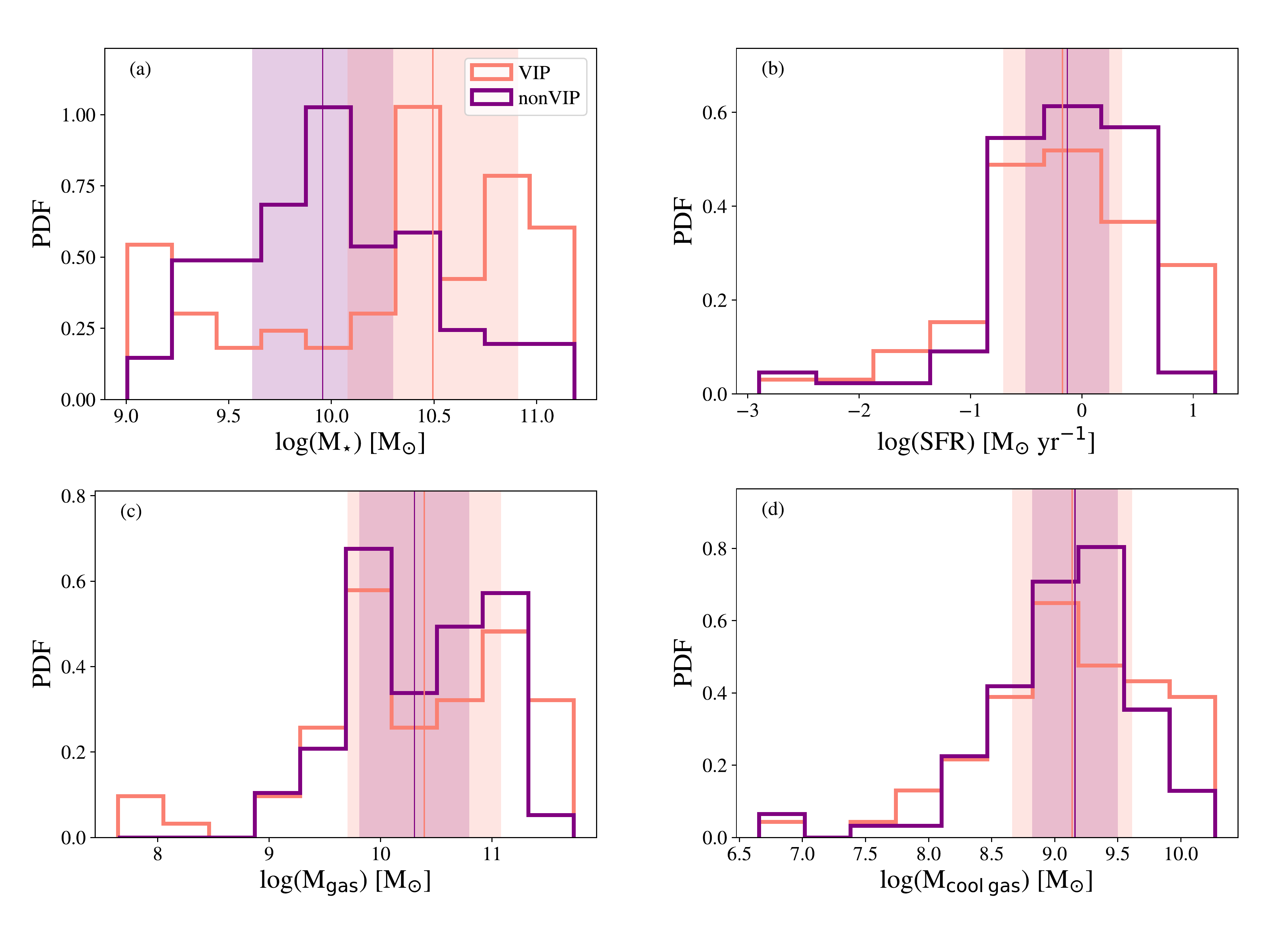}
	\caption{Individual galaxy distributions for the Visually Identified Pairs (VIP; salmon) and the non-visually identified pairs (nonVIP; purple). The panels show: (a) stellar mass, (b) star formation rate, (c) total gas mass, and (d) the star forming gas. The vertical coloured lines correspond to the medians of each sample, and colored rectangles indicate the range within $\pm$ one median absolute deviation from the median.}
	\label{fig:dists_s}
\end{minipage}
\end{figure*}

\subsubsection{Sample Selection Summary}
In Table \ref{table:acnt}, we provide a full account of the results of our various classification schema. Some pairs were manually removed from the sample. This includes systems with multiple prominent subhaloes, which comprised only $\sim$5 per cent of the parent pairs sample. Pairs were also discounted if they have only been in the same group for less than 1 Gyr ($\sim$22 per cent). These are exclusively subhaloes determined not to be interacting, based on the aforementioned criteria. There are a small number of subhaloes ($\sim$2 per cent) which appear to be interacting based on their morphologies in the mock images and/or their trajectories, but were not orbiting one another. Namely, their orbits appear to be dominated by structures outside the FoF group. The majority ($\sim$48 per cent) of our parent sample are not interacting and are not visually identified as mergers. However, there is a small fraction ($\sim$4 per cent) of the non-interacting sample which were misidentified as mergers. Pairs which were visually identified as mergers (\S \ref{sec:viz}) and were found to be interacting (\S \ref{sec:int}) are hereafter referred to as Visually Identified Pairs, or VIP. The nonVIP, then, are those pairs which are interacting, but were not selected visually. The majority of interacting pairs are nonVIP. This is because many of them lack the prominent tidal features or disrupted morphologies that are typically used to identify mergers (\S \ref{sec:viz}; see also Appendix \ref{sec:appb} for visual examples).

\section{Results and Discussion}
\label{sec:results}
\subsection{The VIP and nonVIP Samples}
\label{sec:vnvsamp}
With the sample of pairs defined, and delineated into VIP and nonVIP, we investigate the bulk and individual properties of the interacting pairs at the present day, and at the time of their last pericentres. Among other properties discussed here, we will show that the VIP are not only more massive, but they have undergone a close passage more recently than the nonVIP.
\subsubsection{Present-day ($z=0$) Properties}
\label{sec:prop_z0}
Panel (a) of Figure \ref{fig:dists_bulk} shows that the distribution in stellar mass ratios, which peak at unity and peter out toward larger mass ratios for both samples, with the VIP having slightly larger median stellar mass ratio. Though we do see this trend with the median values, the VIP and nonVIP stellar mass ratios are not distinct distributions: a two-sided KS test indicates ($p\approx0.3$) that these are drawn from the same sample. Panel (b) shows the present-day ($z=0$) 3D separation, with the VIP having separations shifted to smaller values (here, the two-sided KS test indicates that VIP and nonVIP relative separations are drawn from distinct distributions: $p\approx2\times10^{-3}$). Interacting pairs at wider projected separations may be overlooked in preparing samples of merging galaxies. There has, however, been some work which indicates that interacting galaxies may exhibit heightened rates of star formation, even with separations as large as 150 kpc \citep[e.g.][]{patton2013}. In panel (c), we present the relative velocity distributions (i.e. the difference between the subhalo velocities); the VIP and nonVIP samples attain low velocities, facilitating their interaction and eventual merging. The VIP do appear to be moving faster on average than the nonVIP ($p\approx0.05$), which hints at their local dynamics. That is, the relative velocities of these galaxies may be affected by their environment (\S \ref{sec:env}), despite our efforts to avoid this using our FoF group mass cuts. Finally, panel (d) displays the VIP and nonVIP FoF group dark matter mass distributions; the VIP inhabit slightly more massive FoF haloes ($p\approx4\times10^{-4}$). 

In Figure \ref{fig:dists_s}, panel (a) shows that VIP galaxies tend to have higher stellar masses than the nonVIP ($p\approx1.3\times10^{-6}$). Because visual classification is based on tidal disruptions in the stellar material, we might predict that VIP galaxies to have a higher stellar mass (i.e., more stars to disrupt) on average. In fact, we do find that the VIP median stellar mass is about one-half dex greater than the nonVIP, consistent with the findings in panel (d) of Figure \ref{fig:dists_bulk}. Panel (b) shows that the present-day star formation rate (SFR) is relatively consistent for both samples (this is confirmed by a two-sided KS test with $p\approx0.33$). Panels (c) and (d) show the total gas mass and the cool (star forming) gas mass, respectively. The present-day contribution due to cold gas appears to be nearly the same for both samples ($p\approx0.23$). This suggests that the gas reservoir available for star formation is not significantly different for the VIP or nonVIP, consistent with the findings of panel (b).

\begin{figure}
\centering
	\includegraphics[width=\columnwidth]{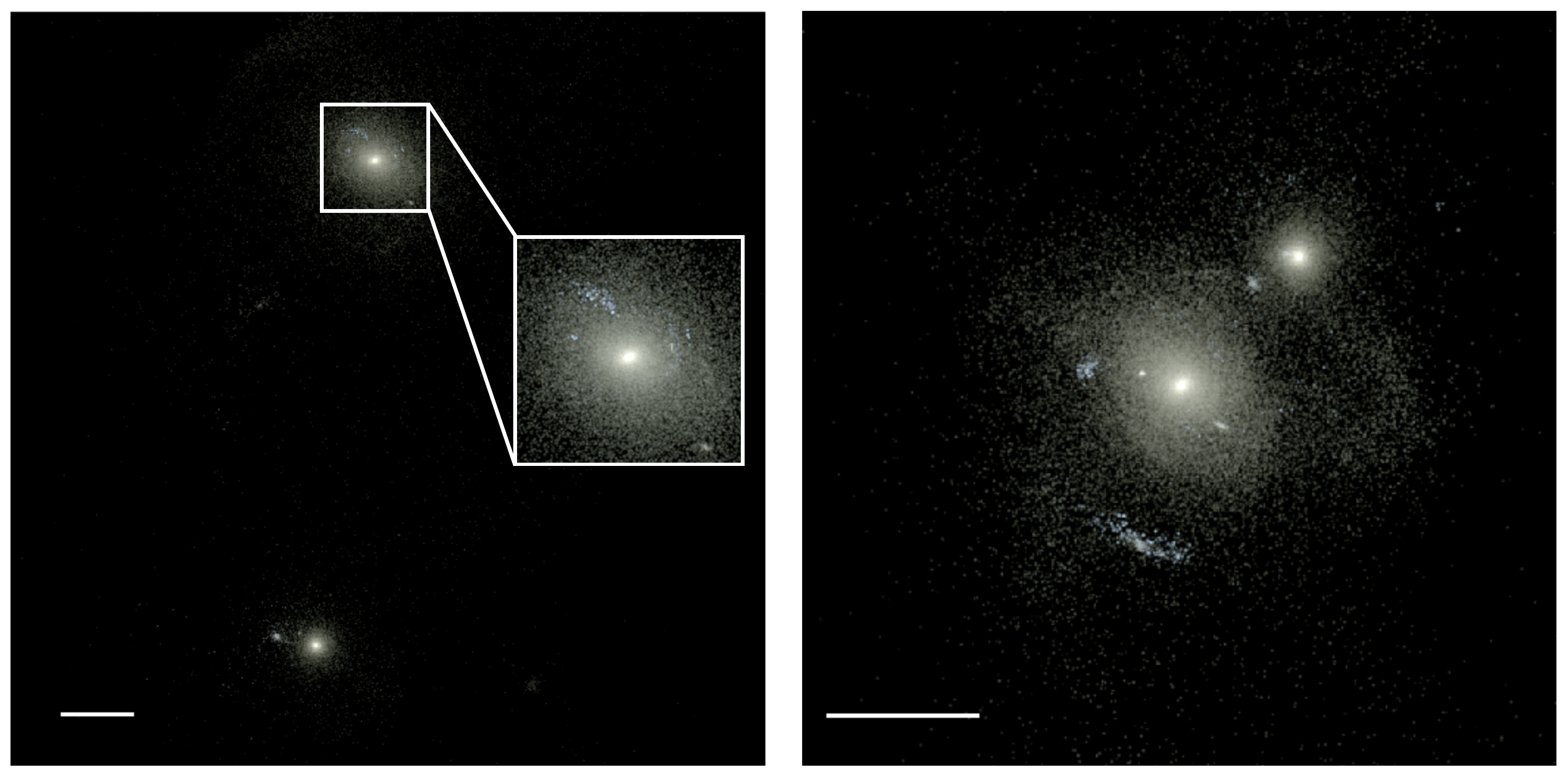}
	\caption{Here we show a (nonVIP) pair at the present day (left) and at its last pericenter (right), which occurred about 1.3 Gyr prior. The stellar mass ratio has stayed at $\sim$3:1 over this period. Bars in the bottom left corner of each image indicate 50 kpc. The inset in the left panel shows an enlarged image of the primary galaxy to highlight its visible structures.}
	\label{fig:z0LP}
\end{figure}

\begin{figure}
\centering
	\includegraphics[trim = 0.6cm 0.57cm 0.3cm 0.35cm, clip, width=\columnwidth]{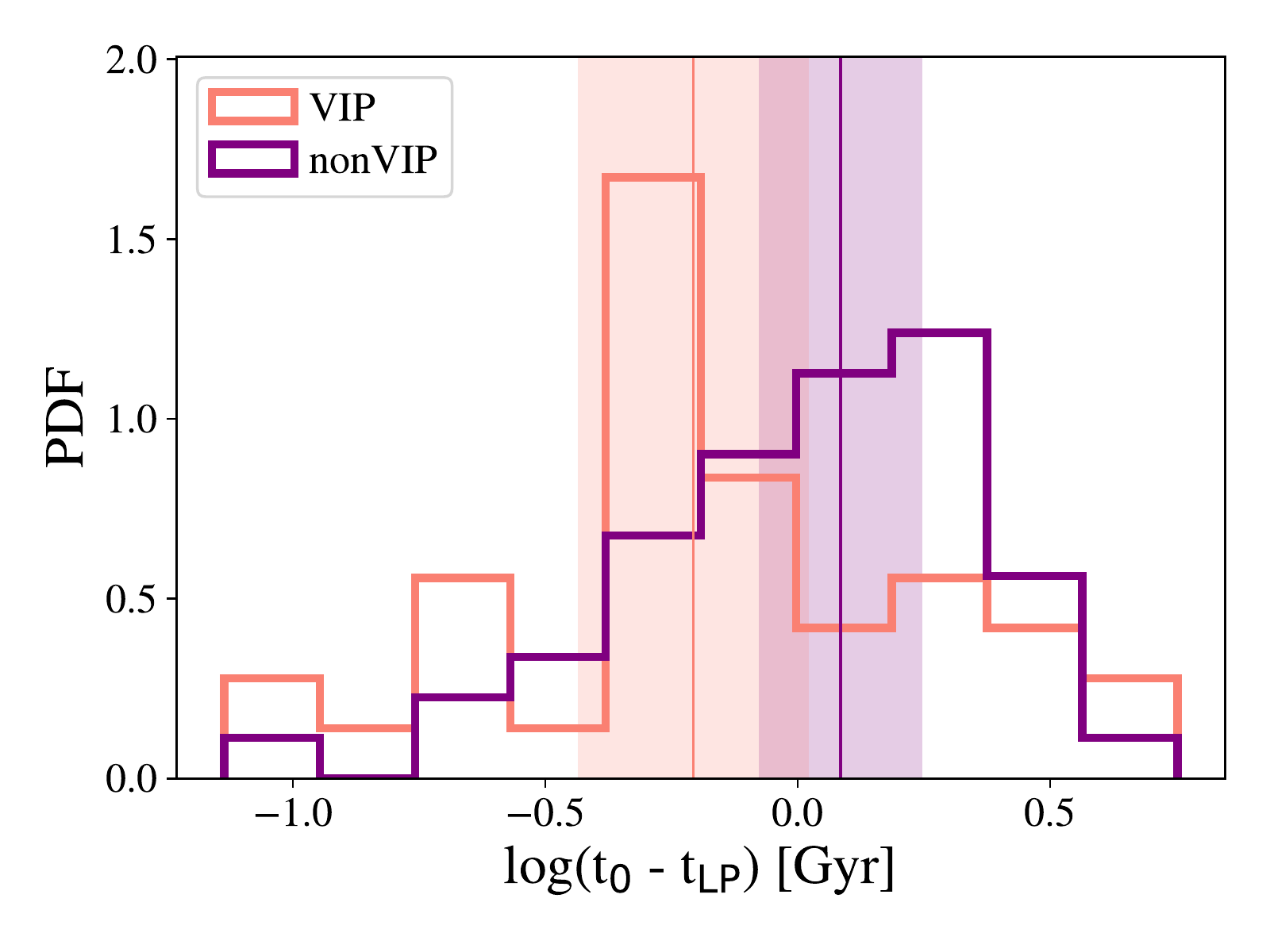}
	\caption{Time since last pericenter (t$_{0}$ - t$_{LP}$) distribution and median values (vertical lines) for the VIP (salmon) and the nonVIP (purple). The VIP have more recently undergone a close passage. Vertical coloured lines show the position of the median, and the coloured rectangles indicate the range within $\pm$ one median absolute deviation from the median.}
	\label{fig:tlast}
\end{figure}

\subsubsection{A Comparative Epoch: Last pericentric (LP) passage}
\label{sec:prop_lp}
We additionally utilise merger trees to study the interacting pairs at the time of their last pericentric passage (LP) -- a local maximum in the strength of their interaction. We note that the interacting pairs do not all reach their respective LP events at the same time, but are at a dynamically similar moment in their histories. In this way, we analyse all interacting pairs at a point in time when the effects of their interaction are at a near a peak. Figure \ref{fig:z0LP} shows a nonVIP galaxy image at the present-day (left), and at its last pericentre (right). At this pair's LP, there is a clear tidal debris field with several star forming regions in the primary galaxy. The LP's span a range of ages relative to the present day of 70 Myr to 5.66 Gyr (Figure \ref{fig:tlast}). The VIP have more recently undergone a close passage than the nonVIP by nearly a factor of two ($p\approx0.01$).

\subsubsection{The Failures of Morphological Identification}
In $\sim$55 per cent of the interacting pairs, it was unclear if an interaction was underway (that is, the nonVIP). There are a number of reasons why morphological identification schema may fail: 
\begin{enumerate}
\item Due to the finite resolution of the simulations, star particles have relatively large masses. This may inhibit our ability to resolve the fine-grained structures within tidal interactions \citep[e.g.,][]{wetz2007}.
\item The stellar material may not be the best indicator of a tidal interaction \citep[discussed as the ``internal properties'' in e.g.,][]{darg2010}. The gas disc has been shown to be as large, if not larger than the stellar disc \citep[e.g.,][]{broeils1997}. Thus, gas discs are much more likely to be perturbed by one another, even in the case of wide pericentric distances. Integral Field Unit surveys \citep[e.g.,][]{croom2012, sanchez2012, bundy2015} of interacting galaxies may be necessary to get a realistic measurement of the local merger rate.
\item The present-day separations (Figure \ref{fig:dists_bulk}) are larger than expected from observationally motivated merger catalogues. What observers assume to be the first passage may, in many cases, be the second \citep[e.g.,][]{patton2013}.
\item If encounters are sufficiently wide, tidal forces may not be strong enough to produce visible (i.e. observable) bridges and tails.
\item If an encounter has occurred within the last Gyr, it is more likely to host obvious tidal features. As time passes, material from the bridge and tails settles back into the discs, and is able to phase-mix with the surrounding material \citep[e.g.,][]{lotz2008, lotz2010}. 
\end{enumerate}

\begin{figure*}
\centering
	\includegraphics[width=0.75\textwidth]{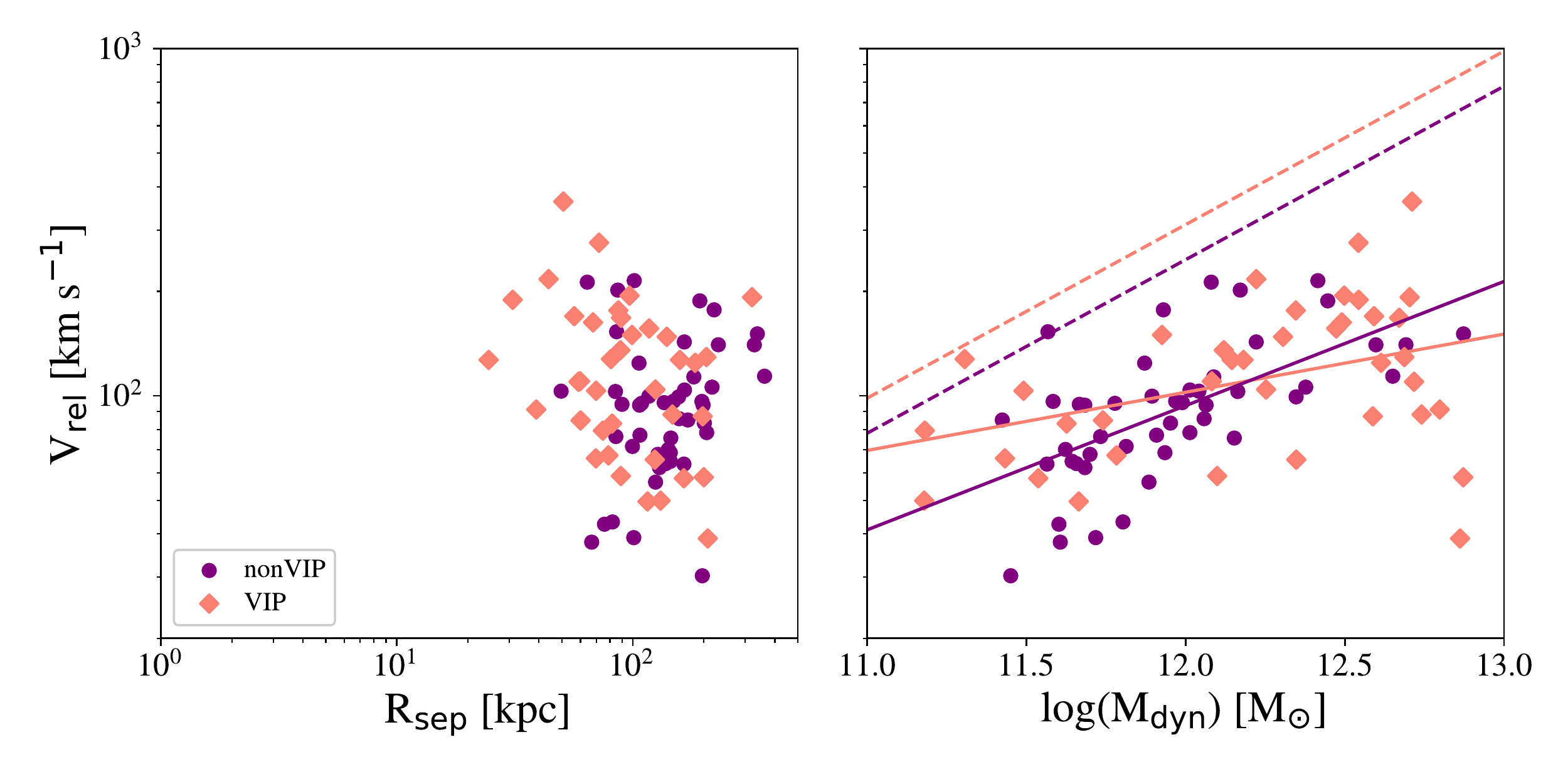}
	\caption{The dynamics of the interacting pairs at $z=0$: relative velocity as a function of pair separation (left) and of dynamical mass (bottom) for the VIP (salmon diamonds) and the nonVIP (purple circles) sample. The dashed lines in the right panel indicate the trend expected from a parabolic trajectory, whilst the solid lines are a fit to the data.}
	\label{fig:dyn_z0}
\end{figure*}

\begin{figure*}
\centering
	\includegraphics[width=0.75\textwidth]{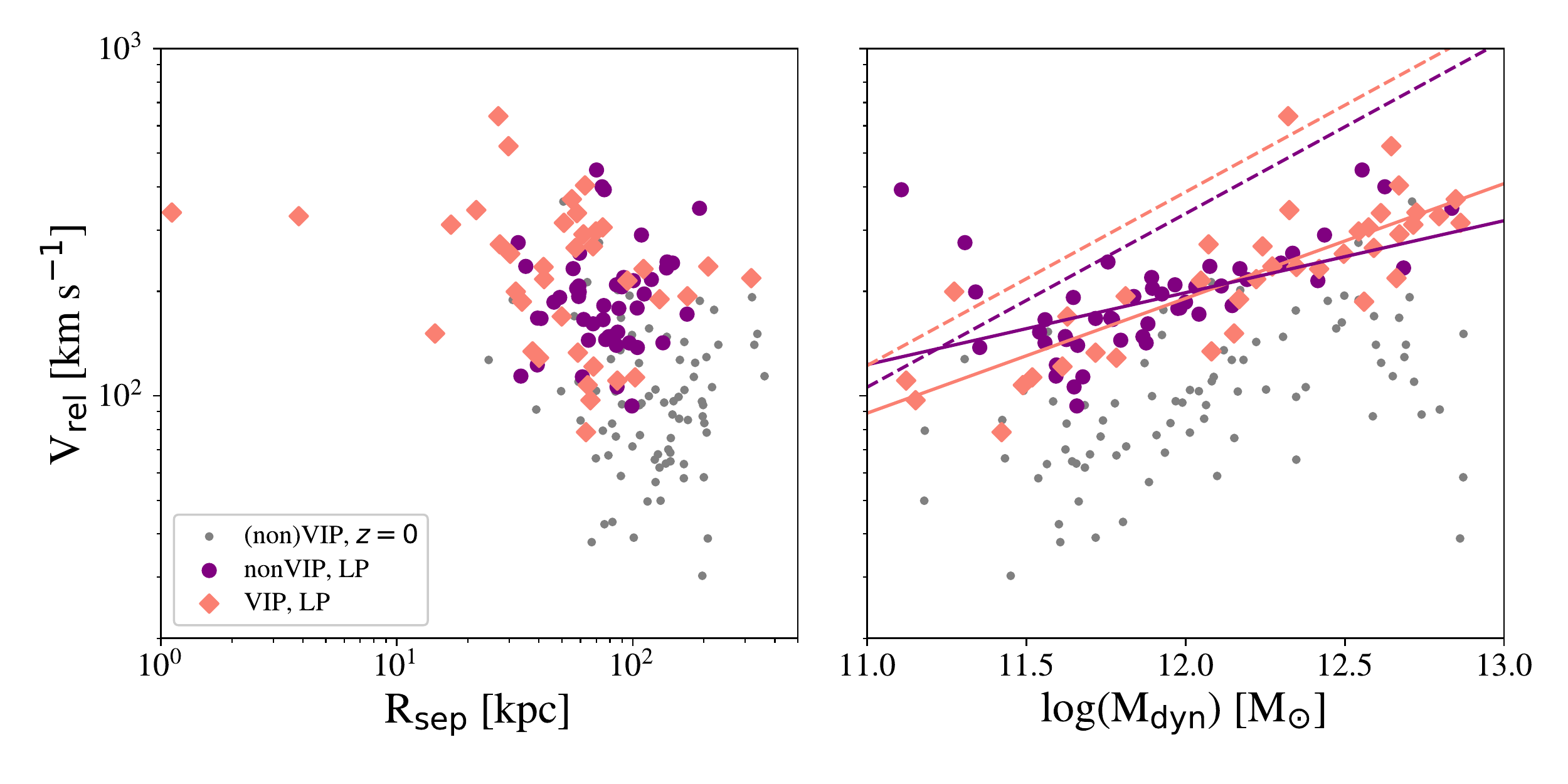}
	\caption{Analogous to Figure \ref{fig:dyn_z0}, but at last pericenter (LP). Values at LP are reported in colour, and in gray points for comparison with $z=0$ values form Figure \ref{fig:dyn_z0}.}
	\label{fig:dyn_LP}
\end{figure*}
\subsection{Galaxy pair dynamics}
\label{sec:dyn_pairs}
To better understand the distinctions between the VIP and nonVIP, we discussed the fundamental physical properties of the pairs in the previous section. Here, we will show that the VIP are closer together and move faster than the nonVIP at both the present day and LP.

Figure \ref{fig:dyn_z0} focuses on the dynamical properties of the VIP (salmon diamonds) and nonVIP (purple circles) at the present day. Consistent with panels (b) and (c) of Figure \ref{fig:dists_bulk}, the left panel of Figure \ref{fig:dyn_z0} shows the VIP and nonVIP are offset from one another: the VIP have smaller 3-dimensional separations, and move with slightly faster relative velocities than the nonVIP. Figure \ref{fig:tlast} shows that the VIP have more recently undergone a pericentric passage. Thus, their separations will naturally be smaller at the present day, and as they are closer to a pericentre, the VIP should have higher velocities than the nonVIP, which are typically closer to an apocentre. The right panel of Figure \ref{fig:dyn_z0} shows a moderate linear trend between the dynamical mass and relative velocity. This is expected if the relative velocity traces the virial velocity, and the orbits are parabolic: $V = \sqrt{2\mathrm{G M}_{\rm dyn} / \mathrm{R}}$ \citep[e.g.,][]{moreno2013}, where the dynamical mass M$_{\rm dyn}$ is defined as the sum of the galaxies' dark matter masses. The ideal trend (i.e., parabolic) is shown for both the VIP (dashed salmon) and nonVIP (dashed purple) samples, using their corresponding median separations at the present day ($\overline{\rm R}_{\rm VIP}$ = 88.8 kpc; $\overline{\rm R}_{\rm nonVIP}$ = 141.4 kpc). The solid lines indicate lines of best fit for each subsample ($m_{\rm nonVIP} = 0.36$ and $m_{\rm VIP} = 0.17$, with standard errors $\sigma_{\rm nonVIP} = 0.067$ and $\sigma_{\rm VIP} = 0.068$). Outliers have a variable impact on the fitted slopes for both the VIP and the nonVIP, ranging from 0.08 to 0.47 dex. Note that their slopes differ from the parabolic case. There is a substantial amount of scatter in these samples, particularly at the high mass end, where a subset of the VIP dip to lower relative velocities. That these galaxies have lower relative velocities than what might be expected based on their dynamical mass may be indicative of their visual identification. 

Figure \ref{fig:dyn_LP} shows the VIP and nonVIP at last pericentre (coloured points), compared with their positions at $z=0$ (gray points). The interacting pairs at LP are significantly closer together, and are moving much faster (left panel) than they are at $z=0$. This is expected: an interacting pair should reach a local maximum in its relative velocity at each pericentric passage \citep[or conversely, should reach a local minimum in its velocity at each apocentre, e.g., Figure 4 of][]{moreno2019}. Similarly, the right panel of Figure \ref{fig:dyn_LP} shows that the the $M_{\rm dyn}-V_{\rm rel}$ relationship is much tighter at LP than at $z=0$ ($m_{\rm nonVIP} = 0.21$ and $m_{\rm VIP} = 0.33$, with standard errors $\sigma_{\rm nonVIP} = 0.05$ and $\sigma_{\rm VIP} = 0.04$). Contrary to the $z=0$ behaviour, the nonVIP appear to have greater scatter, particularly at the low mass end, where a subset achieve higher velocities than their dynamical mass might suggest. The effects of these outliers on the slope is similar to that of the present day population, with a range of 0.02 to 0.33 dex. Moreover, the best fit lines to the data (solid) and the parabolic fiducial curves (dashed) do not agree at last pericenter. This is expected at this epoch, as the orbital elements of the interaction will change rapidly near a close passage. The parabolic trends are elevated from the $z=0$ case, as the median separation values used are $\overline{\rm R}_{\rm VIP} = 57.63$ kpc and $\overline{\rm R}_{\rm nonVIP} = 76.73$ kpc. These curves would be translated down with larger separations. Although the best-fitting lines are still notably different, the VIP slope is now more consistent with the parabolic slopes. 

Deviations from the fiducial parabolic slope hint at the shortcomings of our assumptions regarding galaxy interactions. In particular, this implies that the constant weak gravitational encounters that galaxies experience throughout their evolution impact the orbits in measurable ways. This is made most evident at infall, an epoch discussed in our forthcoming paper (Blumenthal et al. \textit{in prep}.) \\

\subsection{Star formation main sequence}
In Section \ref{sec:vnvsamp}, we presented physical properties such as the stellar mass and star formation rate for the VIP and nonVIP. It is known that these two parameters often trace one another, forming the star formation main sequence.

\begin{figure*}
\centering
\begin{minipage}[t]{0.99\columnwidth}
\vspace{0pt}
\centering
	\includegraphics[trim = 0cm 1.25cm 0cm 0cm, clip,width=\columnwidth]{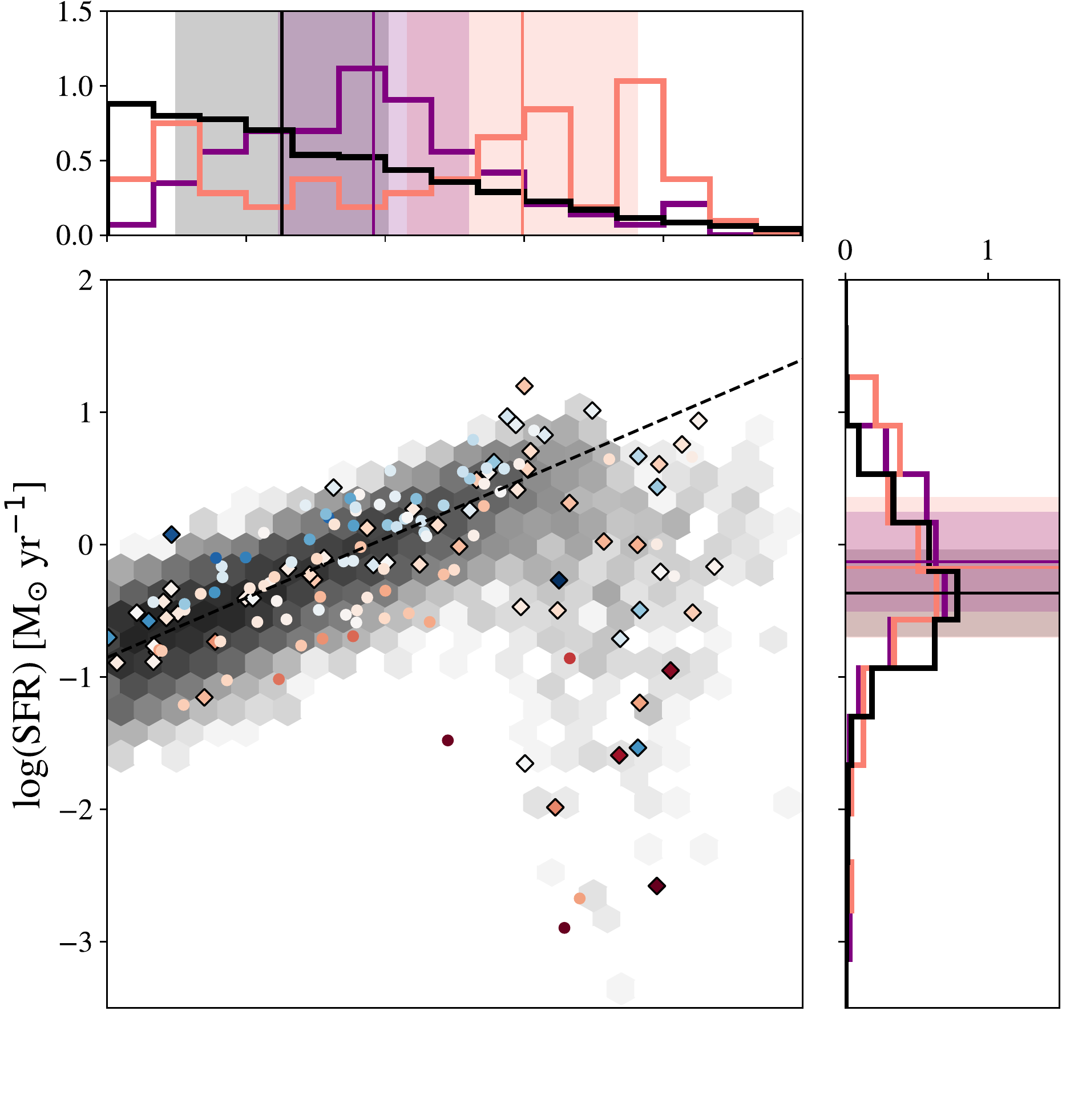}
	\includegraphics[trim = 0cm 1.45cm 0cm 0.78cm, clip, width=\columnwidth]{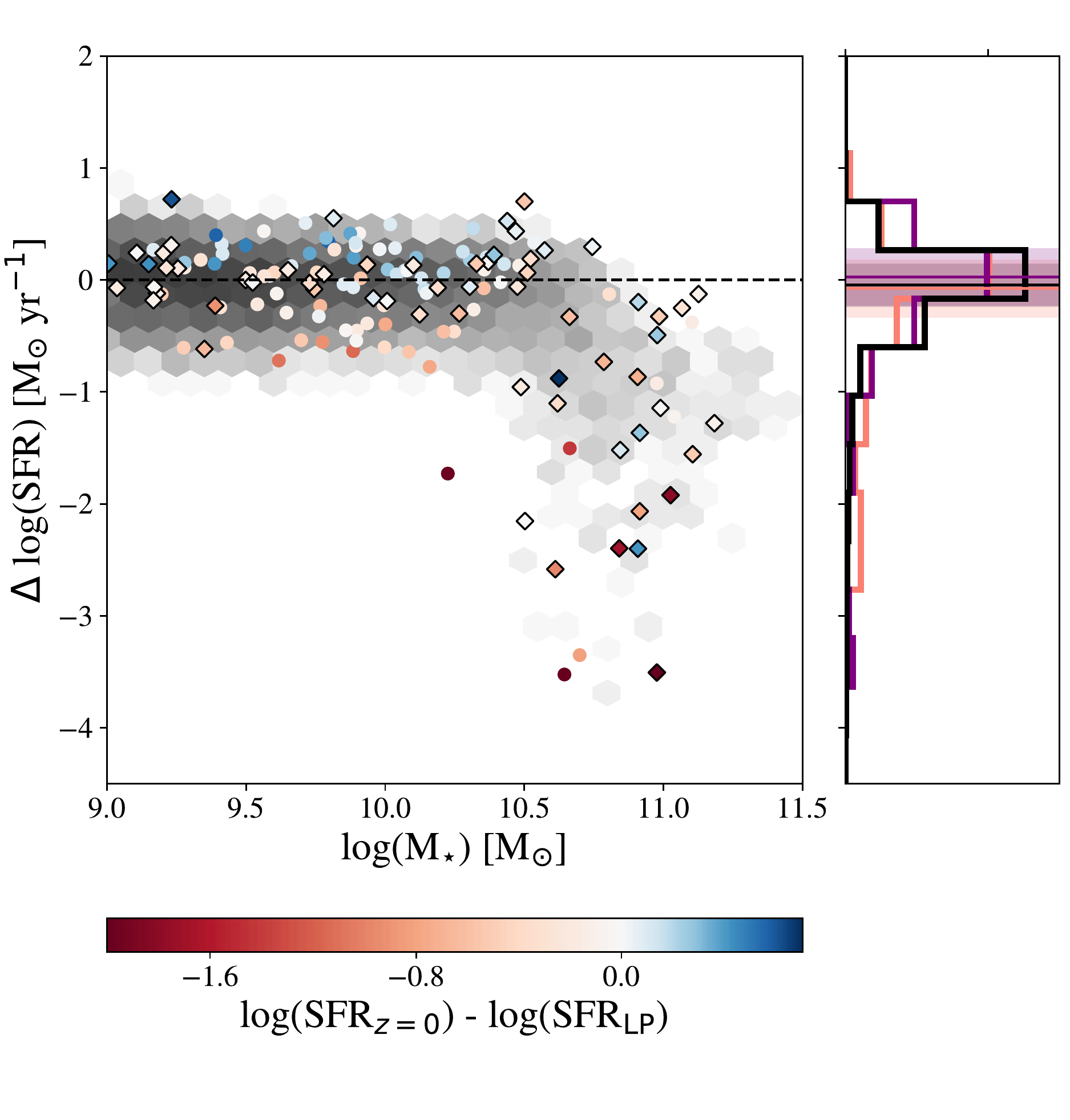}
	\caption{\textit{Top: }The SFMS at the present day. The positions of all TNG100-1 galaxies at $z=0$ which meet the same mass criteria as the interacting pairs are shown in the grayscale hexagons, with the fiducial fit to that SFMS indicated by the black dashed line. The VIP (diamonds, outlined in black) and nonVIP (circles) are coloured by the log in the change of their SFR. \textit{Bottom: }Distance from the SFMS fiducial line, $\Delta$log(SFR) for the same samples as above. Despite the fact that the (non)VIP are interacting pairs, there is no apparent offset \textit{above} the star formation main sequence, though they are offset from the median value of the total TNG100-1 sample (black solid line). The VIP have more scatter in $\Delta$log(SFR), perhaps indicating that their interaction has triggered a dramatic change in morphology, toward compact quiescent spheroids.}
	\label{fig:sfms_z0}
\end{minipage} \hspace{0.04\textwidth}
\begin{minipage}[t]{0.99\columnwidth}
\vspace{0pt}
\centering
	\includegraphics[trim = 0cm 1.25cm 0cm 0cm, clip,width=\columnwidth]{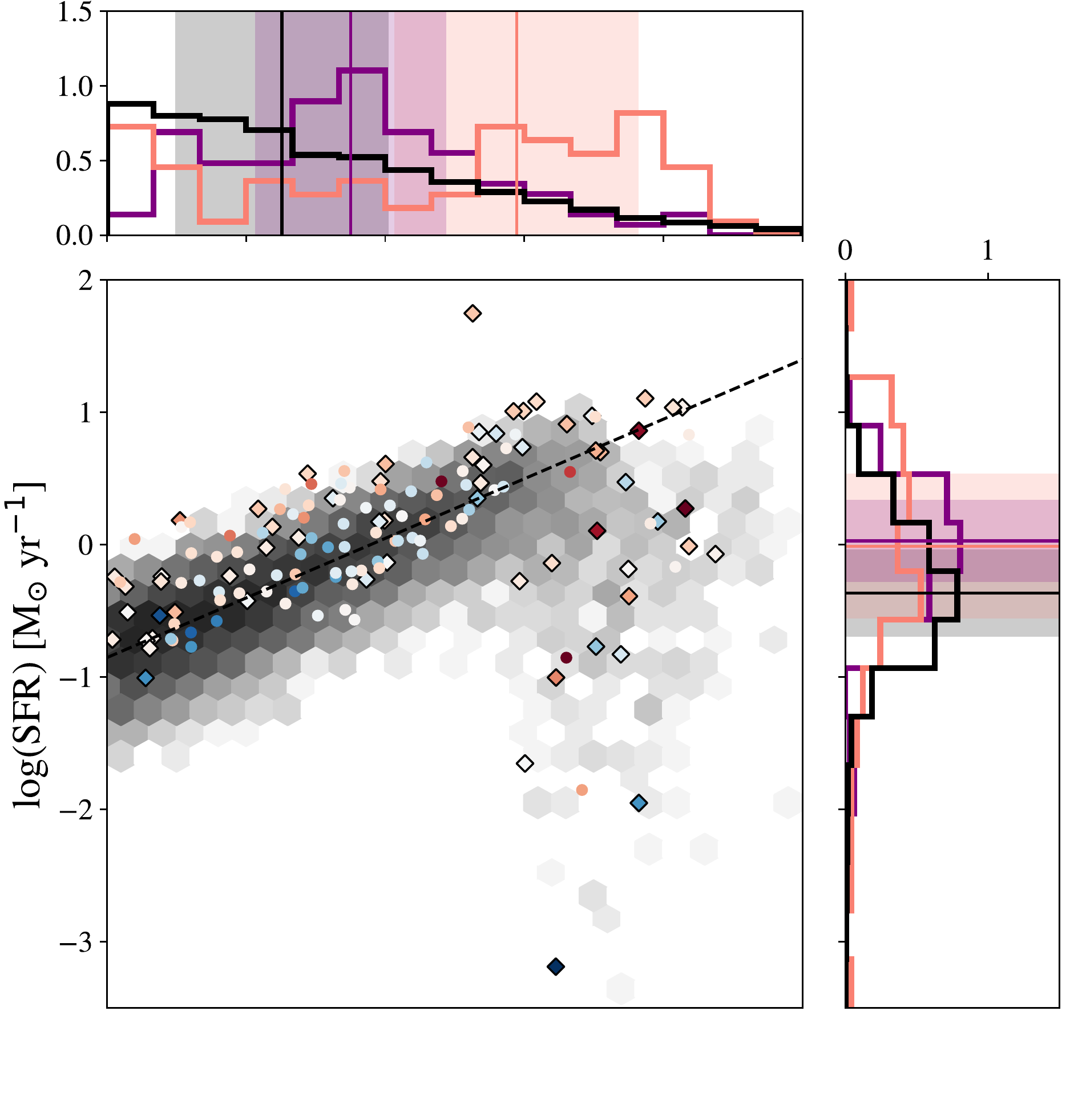}
	\includegraphics[trim = 0cm 1.45cm 0cm 0.78cm, clip, width=\columnwidth]{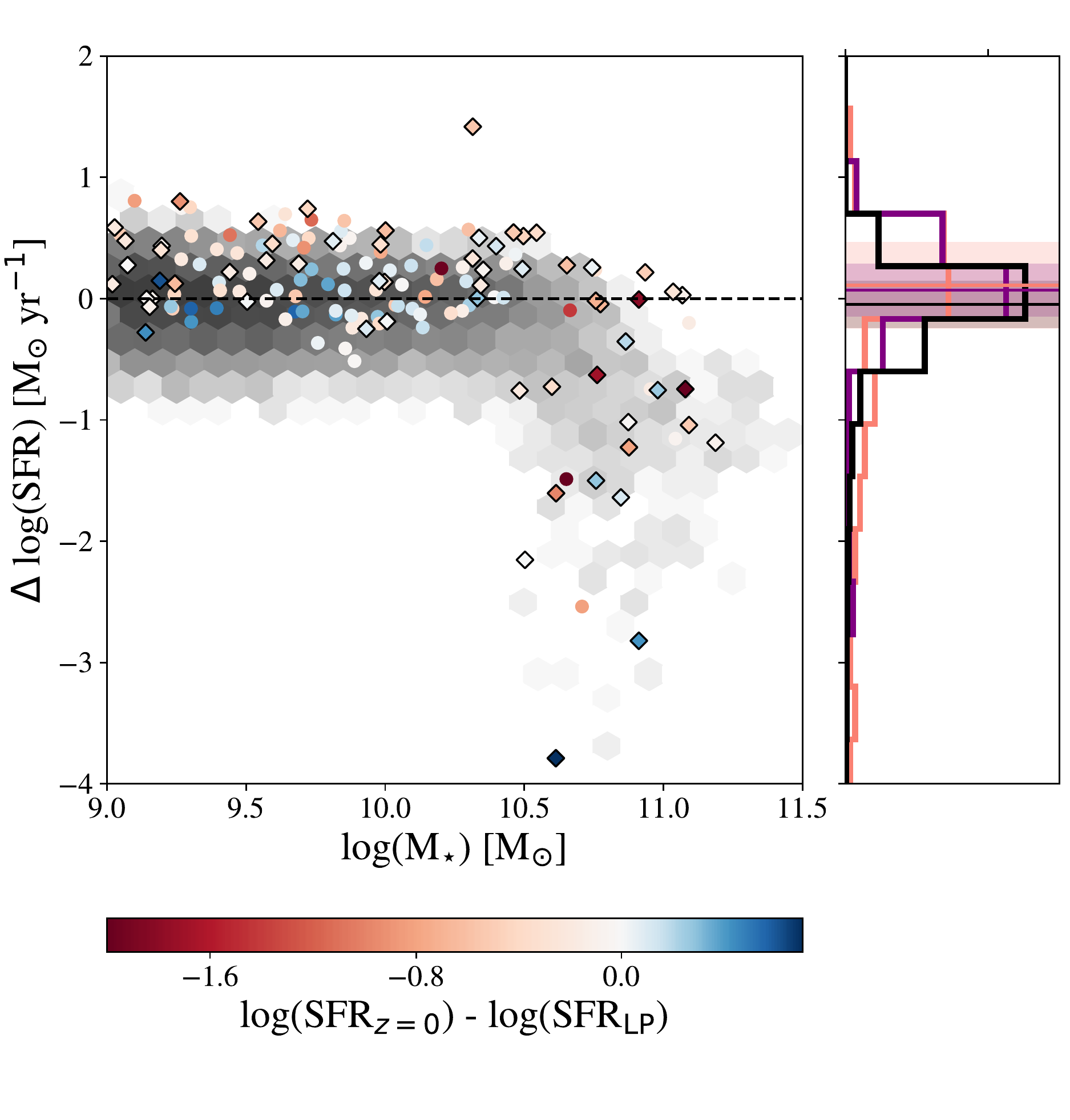}
	\caption{The axes and background galaxies are the same as Figure \ref{fig:sfms_z0}, but shown here are the VIP and nonVIP at their last close passage. Compared to the present day values, the interacting pairs sit slightly higher on the MS, and fewer of them lie in the green valley.}
	\label{fig:sfms_LP}
	\end{minipage}
\end{figure*}

The star formation main sequence \citep[e.g.,][hereafter SFMS]{noeske2007, daddi2007, elbaz2007, salim2007, rodi2011, bluck2016, bluck2019, donnari2019} defines a general trend of all star forming galaxies: the star formation rate is tightly correlated with the stellar mass. That this relationship holds for a wide range of redshifts \citep[e.g., ][]{noeske2007, lee2015}, several orders of magnitude in stellar mass, and a relatively small spread in star formation rate implies that star forming galaxies behave in a self-regulatory manner with a fairly consistent star formation history throughout cosmic time \citep[e.g.,][]{bouche2010, lilly2013}. Outliers above the SFMS (starbursts) are thought to represent an important stage (that is, mergers) in galaxy evolution, though their relative contribution to the star formation density is still debated \citep[e.g.,][]{cox2008, rodi2011, hung2013, brennan2015, willett2015, brennan2017, ellison2018}. Quiescent galaxies lie in a so-called ``red cloud'' below the SFMS, with a ``green valley'' of transitioning galaxies between the two. Merging and interacting galaxies, which themselves are examples of starbursting systems, have been shown to lie above the SFMS \citep[e.g.,][]{puech2014, willett2015}. In particular, \cite{hung2013} show that for $z\sim0.4$ galaxies, distance above the SFMS is correlated with disturbed morphologies. However, other studies \citep[][]{willett2015, brennan2017} are unable to to confirm this morphological dependence. 

Figure \ref{fig:sfms_z0} shows the star formation main sequence for all galaxies (that is, each point is an individual galaxy) at $z=0$ which meet the same mass criteria as the interacting pairs (grayscale hexagons; the black dashed line shows our fiducial SFMS fit), the VIP (diamonds, outlined in black), and the nonVIP (circles). The interacting pairs are coloured by the change in the log of their star formation rates from the present day to LP. These colours enable mapping from $z=0$ to LP, and show how the galaxies have evolved since their last close passage. Blue colours indicate that a galaxy has increased its rate of forming stars since LP, whilst red colours indicate a decrease in SFR since LP. The bottom panel of Figure \ref{fig:sfms_z0} shows the VIP and nonVIP distances, defined as $\Delta$log(SFR) = log(SFR) - log(SFR$|_{\rm{MS}}$), from the fiducial SFMS line as a function of stellar mass. First, we note that the VIP and nonVIP are consistent with the entire set of local TNG100-1 galaxies. The nonVIP appear to exhibit a tight scatter around the SFMS fiducial line, whilst the VIP display a larger spread. The stellar mass appears to increase with increasing FoF group mass, as is expected from abundance matching \citep[e.g.,][]{colin1999, krav1999, krav2004, vale2004, conroy2009, beh2010, guo2010, moster2010}. That there are more VIP at higher stellar and halo masses (see also Figures \ref{fig:dists_bulk} and \ref{fig:dists_s}) may indicate the VIP experience a dramatic change in morphology, perhaps toward compact quiescent spheroids \citep[e.g.,][]{ellison2018}.

Regardless of the orbital geometry, the tidal interaction at a pericentre will invariably draw material from the outskirts of each galaxy toward the center \citep[e.g.,][]{barnes1996, mihos1996, rupke2010, moreno2015, blumenthal2018}. In the case of prograde interactions, a significant amount of gas can be funneled toward the galaxy's nucleus, sparking a burst of star formation \citep[e.g.,][]{alons2000, barnes2004, evans2008, chien2010, moreno2015, lar2016}. Thus, ``observing'' the VIP and nonVIP at the time of their most recent pericentres would naturally push the points in Figure \ref{fig:sfms_z0} up to higher star formation rates. Figure \ref{fig:sfms_LP} shows the star formation main sequence (top) for the interacting pairs at their last pericentres, in addition to the distance from the main sequence fiducial line ($\Delta$log(SFR); bottom). Note that the galaxies which have met the stellar mass threshold at $z=0$ might not achieve this limit at LP. There are only a few galaxies whose masses become unreliable; they are removed from this LP analysis. The colours in both of these panels are the same as in Figure \ref{fig:sfms_z0}. Using these colours, we note that there are some galaxies which appear to move out of the bottom right part of the SFMS (the so-called ``red and dead'' galaxies) between LP and the present day. This may imply that membership in the various regions of the SFMS is fluid: galaxies might undergo periods of starbursts and relative quiescence \citep[e.g.,][]{forbes2014a, forbes2014b}. 

The interacting pairs' shift above the main sequence from the present day to LP cannot be explained by the vertical translation of the SFMS with increasing redshift \citep[e.g.,][]{noeske2007, lee2015} alone. If that were true, the VIP should have systematically lower $\Delta$log(SFR) than the nonVIP in Figure \ref{fig:sfms_LP}. That this is not the case implies the difference in the merger-driven starbursts is mediated by the strength of the interaction, which we have shown is significantly boosted in the VIP. This is supported by the observation that the VIP have higher stellar masses than the nonVIP (Figure \ref{fig:dists_s}), and inhabit FoF groups with more massive dark matter haloes (Figure \ref{fig:dists_bulk}). Studies have shown \citep[e.g.,][]{sobral2011} a connection between the stellar mass, star formation rate, and density of environment. In later sections (\S \ref{sec:env}), we will show that the VIP environment -- partially as measured by the total FoF group mass -- is marginally more dense than that of the nonVIP.

\subsection{Environment}
In the previous sections, we have detailed the intra- and inter-galaxy properties of the interacting pairs. In this section, we describe the external forces acting on these systems through three environmental metrics: the nearest neighbour, the interaction strength, and the FoF group mass.
\label{sec:env}
\begin{figure*}
	\centering
	\begin{minipage}[t]{0.3\textwidth}
		\vspace{0pt}
		\centering
		\includegraphics[width=1.05\columnwidth]{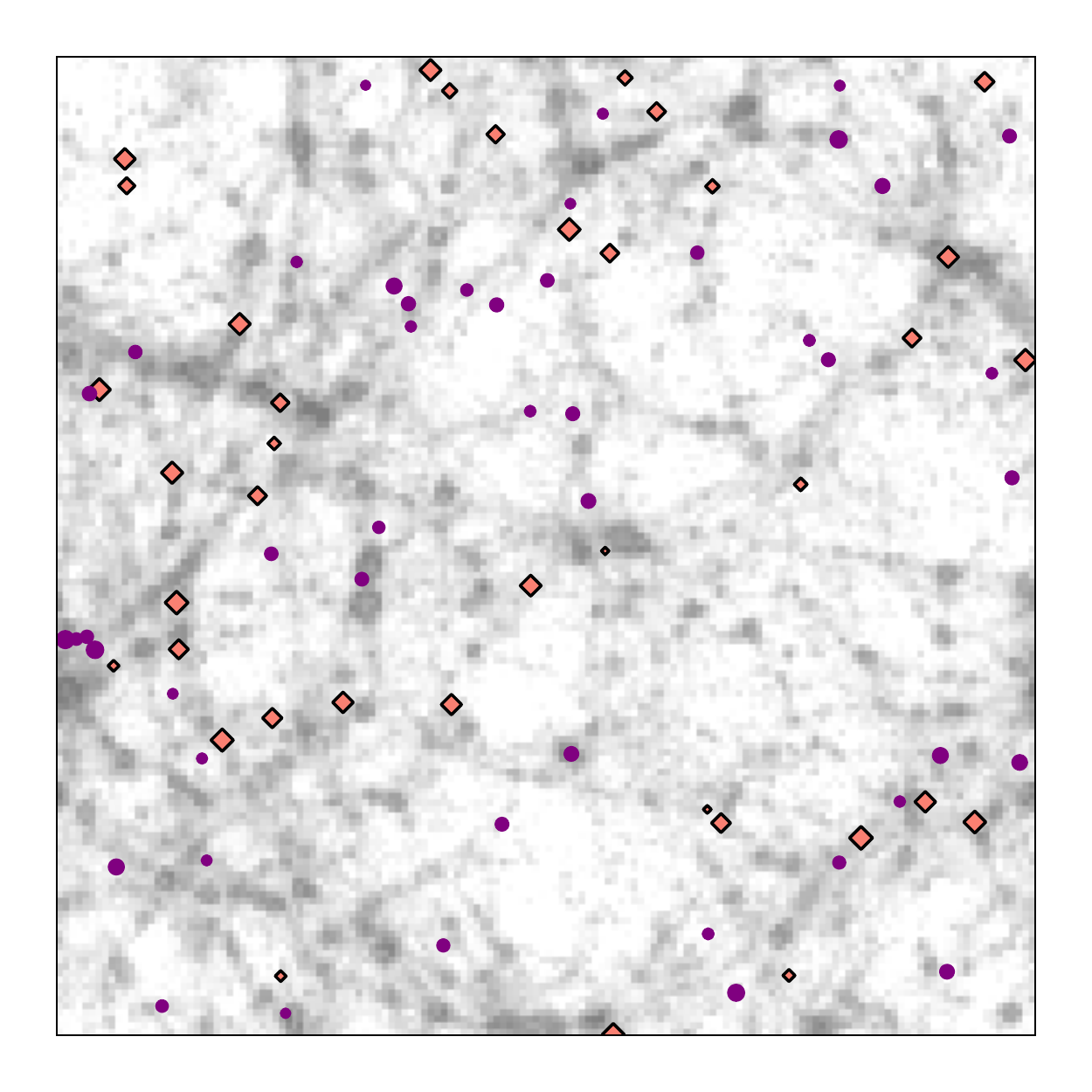}
		\includegraphics[trim = 0.6cm 0.57cm 0.3cm 0.35cm, clip, width=\columnwidth]{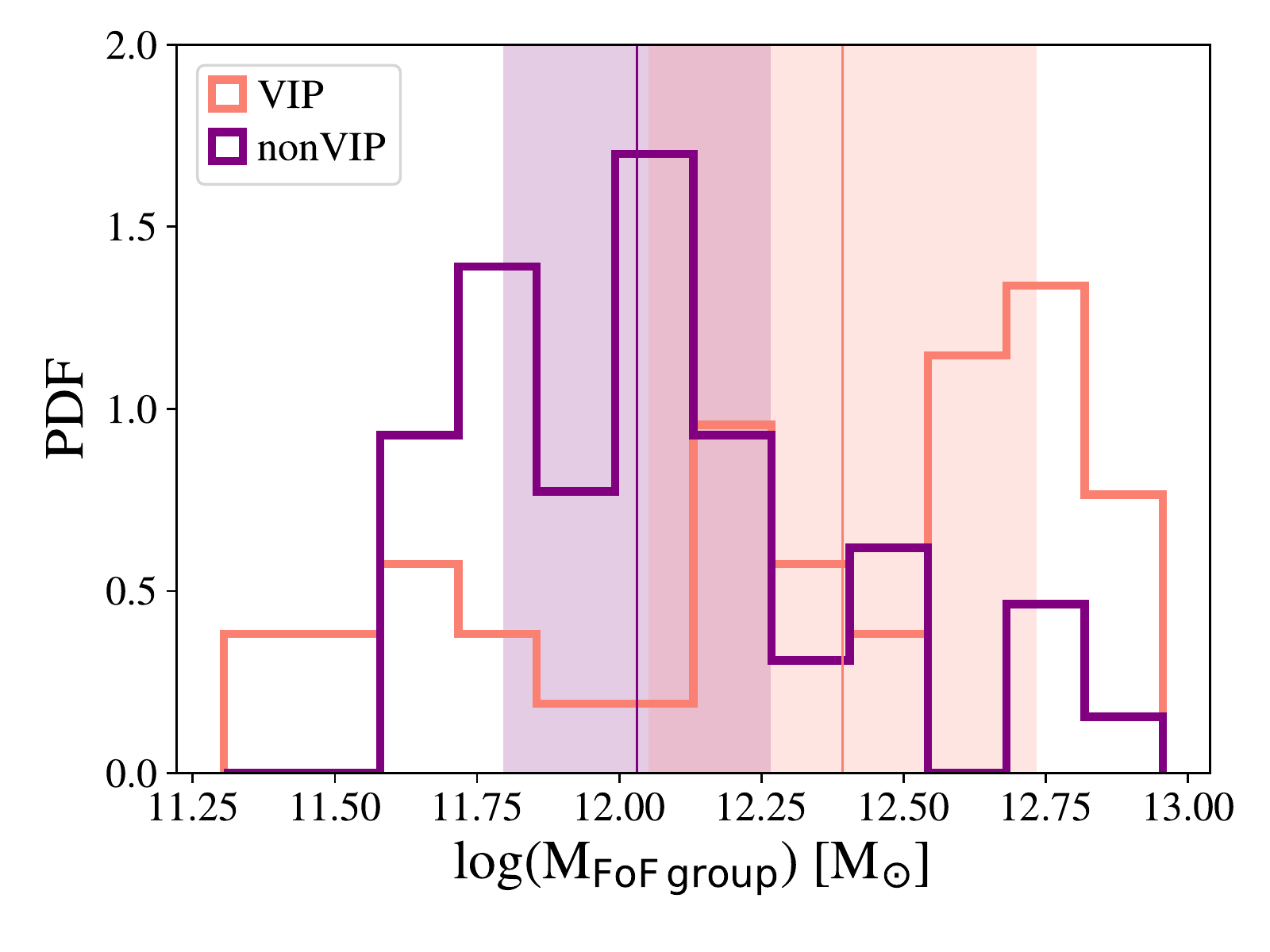}
	\end{minipage} \hspace{0.01\textwidth}
	\begin{minipage}[t]{0.3\textwidth}
		\vspace{0pt}
		\centering
		\includegraphics[width=1.05\columnwidth]{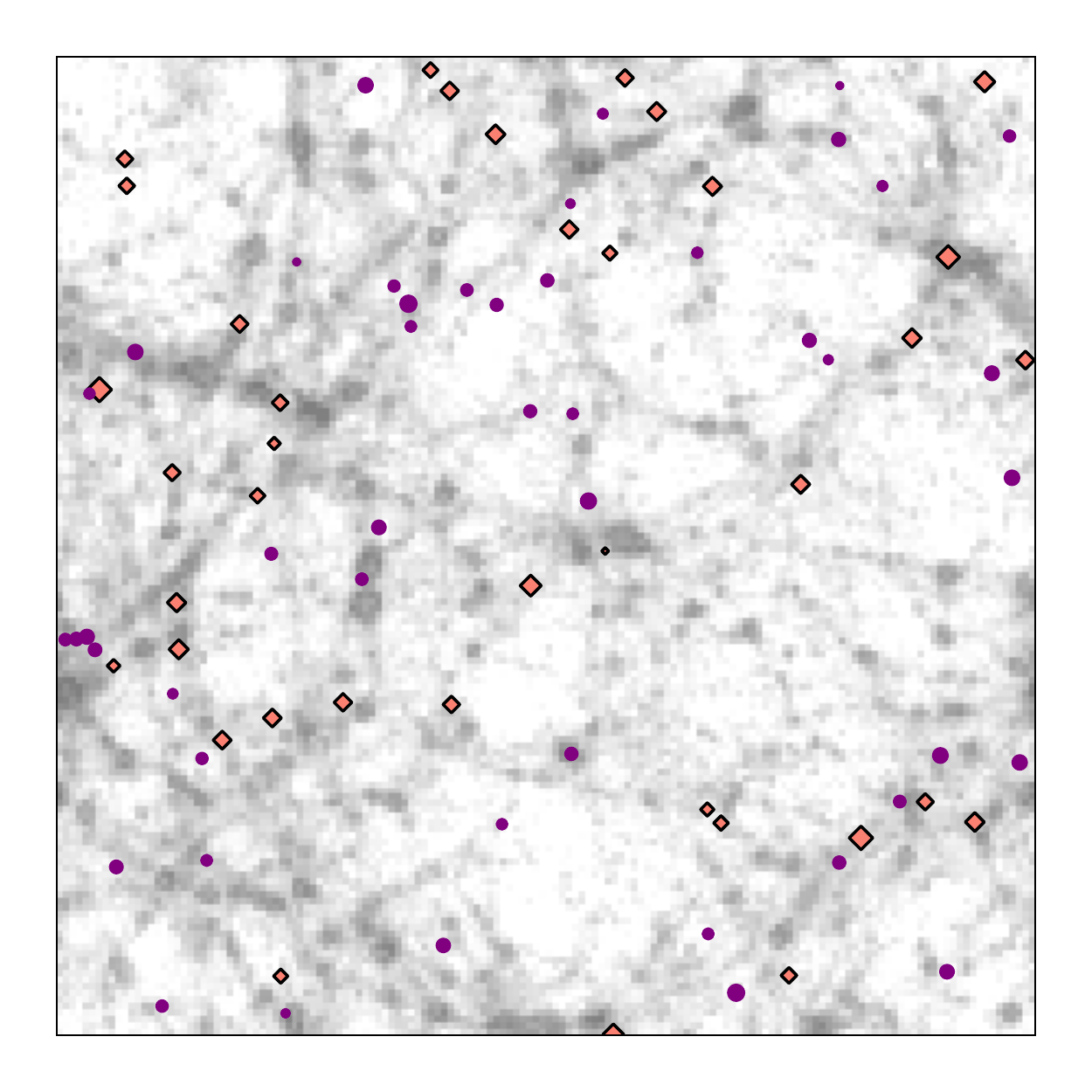}
		\includegraphics[trim = 0.6cm 0.57cm 0.3cm 0.35cm, clip, width=\columnwidth]{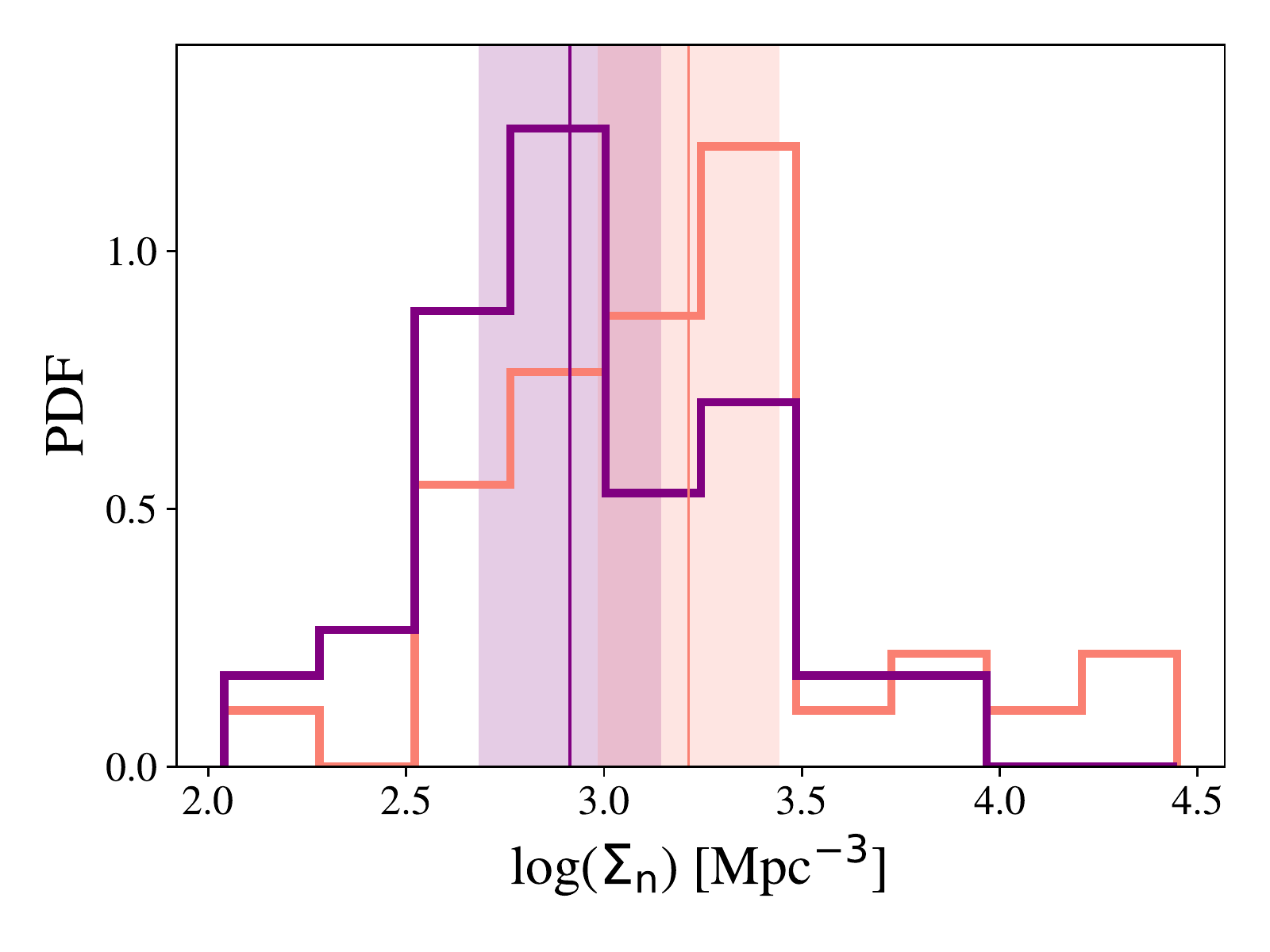}
	\end{minipage} \hspace{0.01\textwidth}
	\begin{minipage}[t]{0.3\textwidth}
		\vspace{0pt}
		\centering
		\includegraphics[width=1.05\columnwidth]{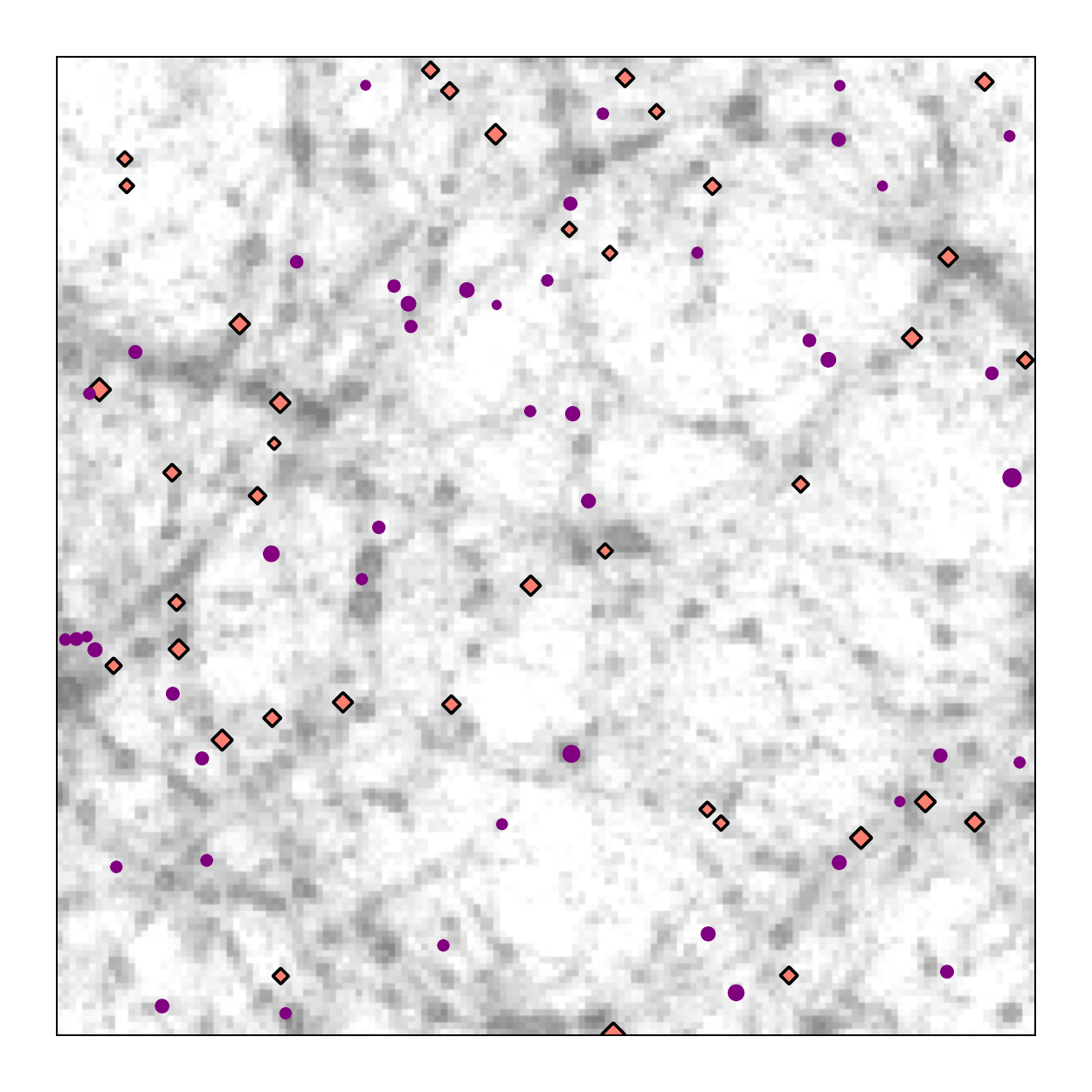}
		\includegraphics[trim = 0.6cm 0.57cm 0.3cm 0.35cm, clip, width=\columnwidth]{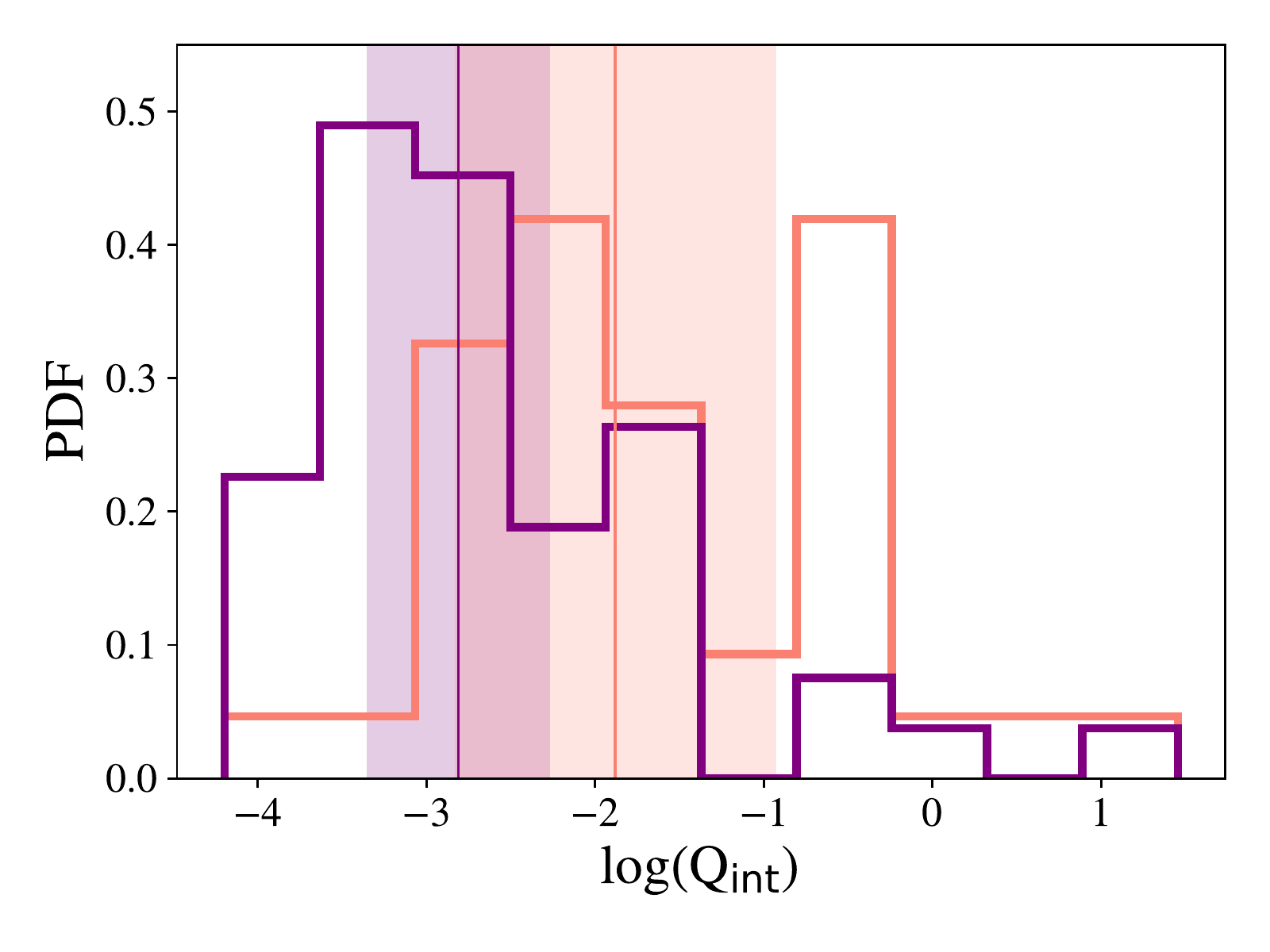}
	\end{minipage}
	\caption{\textit{Top: }Placement of the nonVIP (purple circles) and VIP (salmon diamonds) in the TNG100-1 cosmic web. The point sizes are scaled by the total group mass (M$_{\rm FoF \, group}$, left), nearest neighbour statistic ($\Sigma_{\rm n}$, centre), and the interaction strength ($Q_{\rm int}$, right). \textit{Bottom: }We also show the distributions of the three environmental measures for the VIP (salmon) and nonVIP (purple), as before. Vertical coloured lines show the median values, and the corresponding coloured rectangles indicate the range within $\pm$ one median absolute deviation of the median. The VIP sit in significantly more massive haloes, in denser environments and are more affected by their surroundings than the nonVIP.}
	\label{fig:web}
\end{figure*}

There is no universal definition of galactic environment \cite[e.g.,][and references therein]{muldrew2012}. Many studies attempt to compare the various definitions of this fundamental property \citep[e.g.,][]{cooper2005, gallazzi2009, wilman2010, haas2012, shattow2013, fossati2017}. Parameterisations used to characterise the environment include the local number density \citep[e.g.,][]{dressler1980, lewis2002, cooper2005, shattow2013}, measurements of galaxy clustering \citep[e.g.,][]{skibba2013, skibba2015, gun2018}, and placement within cosmic structures \citep[e.g.,][]{yang2007, darvish2014, kuutma2017, liao2019}.

The top panels of Figure \ref{fig:web} show the placement of the VIP (salmon diamonds) and nonVIP (purple circles) samples in the TNG100-1 cosmic web, as traced by all subhaloes in the $z=0$ slice. Qualitatively, the VIP typically lie in denser regions than the nonVIP (which seem to mostly occupy voids). The marker sizes in each of the three panels scale linearly with the log of three independent measures of environment: the total FoF group mass (left, \S\ref{sec:fofm}), the n$^{\rm th}$ nearest neighbour, $\Sigma_{\rm n}$ \citep[e.g.,][middle, \S\ref{sec:sigman}]{dressler1980, lewis2002}, and the interaction strength, $Q_{\rm int}$ \citep[e.g.,][right, \S\ref{sec:qint}]{verley2007}. The structure of the cosmic web is defined by the location of haloes containing individual, groups and clusters of galaxies. Thus, the mass of a halo is indicative of its placement within this structure. The nearest neighbour statistic measures environment based on the number density of nearby galaxies, regardless of mass. On the other hand, the interaction strength measures the balance of external tidal forces from all galaxies within an aperture with the binding force of the galaxy. Whilst it can be difficult to discern any trends from the cosmic web panels, the bottom panels of Figure \ref{fig:web} show the subsequent distributions for each of these environmental metrics. 

\subsubsection{FoF Group Mass}
\label{sec:fofm}
The bottom right panel of Figure \ref{fig:web} shows the distribution of FoF group total masses for the VIP (salmon) and nonVIP (purple). This indicates that the VIP sit in preferentially more massive haloes ($p\approx4\times10^{-4}$), surpassing the nonVIP by nearly half an order of magnitude. This is consistent with the fact that most massive haloes are likely to sit in nodes or at intersections of filamentary structures \citep[e.g.,][and sources therein]{bond1996, joachimi2015}. 

\subsubsection{n$^{th}$ Nearest Neighbour}
\label{sec:sigman}
The n$^{\rm th}$ nearest neighbour statistic is a number density measurement that uses the distance to the n$^{\rm th}$ nearest neighbor, r$_{\rm n}$, to define the volume. That is, 
\begin{equation}
	\Sigma_{\rm n} = \frac{\rm n - 1}{\frac{4}{3} \pi \mathrm{r}_{\rm n}^{3}}
\end{equation}
where the numerator n$-1$ is used to discount the central or, primary galaxy. (Note that here we employ a three-dimensional version of what is typically used by observers.) Thus, centrals with larger $\Sigma_{\rm n}$ sit in denser environments. For the purposes of this paper, we adopt n$=5$. The bottom left panel of Figure \ref{fig:web} shows that the VIP lie in preferentially denser environments than the nonVIP ($p\approx0.07$). 

\subsubsection{Interaction Strength}
\label{sec:qint}
One major drawback of the $\Sigma_{\rm n}$ measure is that it does not account for the mass of neighbouring galaxies. The interaction strength, $Q_{\rm int}$, thus serves a useful counterpoint to  $\Sigma_{\rm n}$ in its careful accounting of the tidal effect of nearby galaxies. \cite{verley2007} defined the interaction strength as the ratio of the cumulative tidal forces tugging on the galaxy from all neighbouring galaxies within a set aperture, and the binding force keeping the central together: 
\begin{equation}
	Q_{\rm int} \equiv \frac{F_{\rm tidal}}{F_{\rm bind}} \>\> , \>\>
	F_{\rm tidal} = \frac{M_{\rm n} D_{\rm c}}{R_{\rm nc}^{3}} \>\> , \>\>
	F_{\rm bind} = \frac{M_{\rm c}}{D_{\rm c}^2}
\end{equation}
where $M_{\rm n}$ is the mass of the neighbor, $R_{\rm nc}$ is the distance from the central galaxy to that neighbor, $M_{\rm c}$ is the mass of the central, and $D_{\rm c}$ is the diameter of the central. Following observational studies (for which this metric was developed), we take all masses to be the total mass within twice the stellar half-mass radius, and the diameter of the central galaxy which corresponds to that mass (that is, four times the stellar half-mass radius). This value is calculated in a number of different ways in \cite{verley2007}, including using a fixed and infinite aperture (that is, all galaxies within a fixed volume). They find that there was very little difference between the two, as distant galaxies will contribute only a small amount to the tidal field of the central. To accommodate the large present-day separations of some of our interacting pairs, we use an aperture of 5 Mpc (Figure \ref{fig:web}, bottom right). The VIP are affected by the tidal effects of their neighbours nearly ten times as much as the nonVIP ($p\approx7\times10^{-4}$).

\subsubsection{The Effects of Environment}
In the previous subsections, we demonstrated that the VIP belong to more massive FoF haloes, sit in denser environments, and are more affected by interactions with their neighbours than the nonVIP. Here we disentangle the effects of mass and environment and show that although the (more massive) VIP are in systematically more dense environments, there is no statistically significant difference between the $\Delta$log(SFR) of the nonVIP and VIP, when controlling for stellar mass. 

\begin{figure}
\includegraphics[trim = 0.35cm 0.5cm 0.5cm 0.5cm, clip, width=\columnwidth]{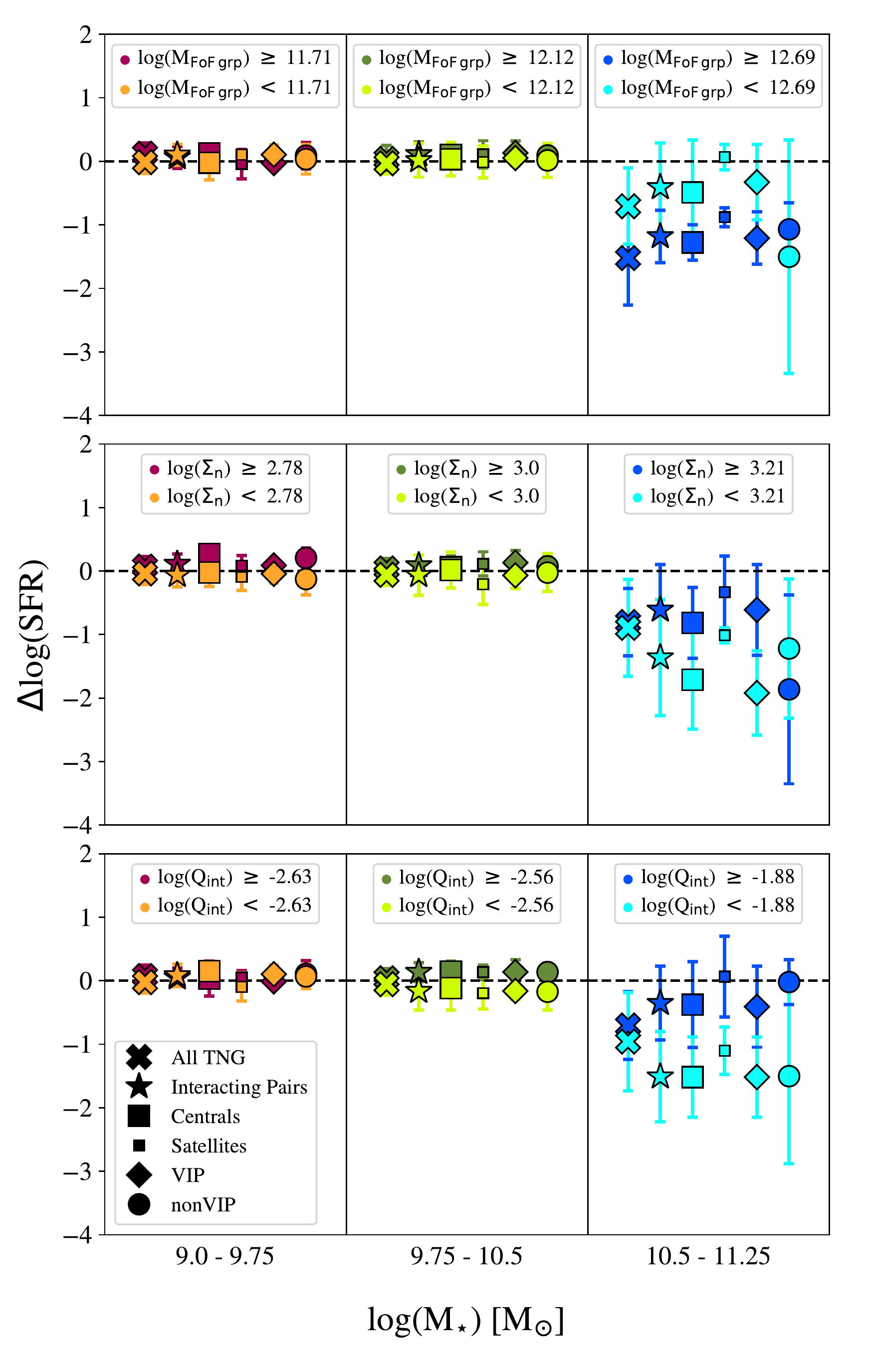}%
	\caption{We separate the stellar mass into three bins, and further split the subsample into two environmental bins based on the median environmental measure in that mass bin. Stars indicate the median value of $\Delta$log(SFR) for all interacting pairs within each mass bin with error bars that correspond to the median absolute deviation. Markers represent the the median $\Delta$log(SFR) for the galaxies within that mass bin. This is shown for the FoF group mass (top), $\Sigma_{\rm n}$ (middle), and Q$_{\rm int}$ (bottom). X's represent the background distribution of all TNG100-1 galaxies. Stars indicate the sample of interacting pairs (that is, the VIP and nonVIP together). As before, the nonVIP and VIP are indicated by circles and outlined diamonds, respectively. Satellites and centrals are also shown, and are distinguished by symbol size: small and large, respectively.}
	\label{fig:envsfr}
\end{figure}

Figure \ref{fig:envsfr} shows the $\Delta$log(SFR) as a function of stellar mass. The stellar mass distribution is split into three bins: 9.0 $\leq$ log(M$_{\star}$) $<$ 9.75, 9.75 $\leq$ log(M$_{\star}$) $<$ 10.5, and log(M$_{\star}$) $\geq$ 10.5. These mass increments were chosen to separately analyse ``normal'' star forming galaxies (log(M$_{\star}$) $<$ 10.5) from those which have begun to dip below the main sequence (log(M$_{\star}$) $\geq$ 10.5). Environment is considered independently within each of these bins. Different colours indicate the galaxies which sit in a relatively low (light colours) or high (dark colours) density environment. The median environmental measure in each mass bin is used to delineate between low and high densities. We present this in Figure \ref{fig:envsfr} for the M$_{\rm FoF \, group}$ (top), $\Sigma_{\rm n}$ (middle) and Q$_{\rm int}$ (bottom). As before, the VIP and nonVIP are distinguished by diamond and circle markers, respectively. Centrals and satellites are indicated by the marker size (large and small, respectively). The median $\Delta$log(SFR) values for both environment bins within each mass bin are displayed as stars with error bars indicating the  first and third quartiles (that is, the width of the distribution). For comparison, the median value of $\Delta$log(SFR) of the underlying TNG100-1 distribution is displayed by the X's.

The first two mass bins of all three environmental metrics shown in Figure \ref{fig:envsfr} show no clear trend with environment. That is, not only do they show no distinction between high and low density environments, but they are consistent with the background distribution of all TNG100-1 galaxies (coloured X's). Only in the largest mass bin do we see any significant difference between the low and high density environments across all samples. In the largest mass bin of the top panel (M$_{\rm FoF \, group}$), the less massive FoF groups have systematically higher $\Delta$log(SFR) than the high mass FoF groups, as these are likely quenched or are in the process of quenching. It should be noted however that in this panel, the interacting pair sample (and its individual components) are consistent with the background. 

The middle and bottom panels of Figure \ref{fig:envsfr} indicate that denser environments foster higher star formation rates only within the highest mass bin. Whereas before, the interacting pairs behaved similarly to the background TNG100-1 galaxies, in the $\Sigma_{\rm n}$ and Q$_{\rm int}$ panels, the interacting pairs diverge significantly from the background TNG100-1 galaxies. Further, that the environmental dependence of $\Delta$log(SFR) only becomes appreciable at higher masses -- when AGN activity and quenching begin to dominate a galaxy's evolution -- implies environment plays a larger role in suppressing quenching than it does in boosting star formation. 

Though there is no clear distinction between the VIP and nonVIP at any mass bin (except for the highest mass bin of Q$_{\rm int}$), the satellites and centrals appear to have divergent evolutionary pathways. The centrals dip low in $\Delta$log(SFR) at high masses whereas the satellites are only moderately affected. This implies that centrals are likely to quench before their satellites. It may be that the evolution of satellites is more sensitive to environment, whilst the evolution of centrals is depends more strongly upon mass; perhaps an example of the interplay between ``environment quenching'' and ``mass quenching'' \citep[e.g.,][]{peng2010, bluck2016, bluck2019}. Thus, the relative importance of environment and stellar mass  depends upon which component of the interaction is the subject of inquiry.

\section{Conclusions}
\label{sec:conclusions}
In this paper, we identify a set of paired galaxies from the $z=0$ snapshot of the TNG100-1 simulation of IllustrisTNG \citep{marinacci2018b, naiman2018, nelson2018, pill2018b, springel2018}. We generate ideal mock SDSS $g$-band images of all pairs and visually classify each as interacting or not interacting. We then confirm using the information from the Sublink merger tree, and find that of the interacting pairs, we correctly identify 38 (the Visually Identified Pairs, or VIP) and miss 47 (the Non-Visually Identified Pairs, or nonVIP). Our analysis includes a detailed study of the interacting pairs' present day properties, as well as their properties at their respective last pericentres. \\

Our primary findings are as follows:
\begin{enumerate}
	\item Stellar morphologies are not ideal for identifying interactions as the visibility of stellar tidal features, which is in part dependent upon the environment and the time since the last close passage.
	\item Using the merger trees, we trace the interacting pairs back to their time of last pericentric (LP) passage.
	\item The VIP have more recently undergone a close passage than the nonVIP by about a factor of two. As a result, their tidal features are easier to observe. Merger classifications are thus biased toward recent interactions.
	\item Compared with the nonVIP, the VIP sit in very different environments. The VIP are: in groups which are nearly 2.5 times as massive; in nearly twice as dense surroundings; and are affected by interactions with their neighbours by nearly an order of magnitude more than the nonVIP. Classification schema based on stellar morphologies are biased toward dense environments.
	\item Though the VIP sit in distinct environments from the nonVIP, the visibility of a pair does not depend strongly on environment, when correcting for stellar mass. 
\end{enumerate}

Care should be taken when interpreting these results within the context of large observational catalogues of galaxy pairs \citep[e.g.,][]{ellison2008, ellison2010, patton2016}, as the mass range covered in this work is relatively limited. The roughly 45\% merger recovery rate that we present here should \textit{not} be taken as a completeness correction. This work can only offer a critique of the observational surveys used to derive merger rates \citep[e.g.][]{kartaltepe2007, kitzbichler2008, robo2014}. To fully answer this question would require a realistic mock (that is, not ideal mock as described above) survey of all galaxy pairs, though this is beyond the scope of this work. However, we have illuminated distinct biases inherent in observational galaxy pair catalogs. If these are used to determine the merger rate, the result is likely to be biased toward close pairs (\S\ref{sec:dyn_pairs}), high stellar masses (\S\ref{sec:prop_z0}) that may result in particularly prominent tidal features, and recent pericentric passages (\S\ref{sec:prop_lp}). 

$\>$

In addition to the intrinsic and dynamic properties of galaxies, the production of tidal features depends on the interaction geometry. In a forthcoming paper (Blumenthal et al. \textit{in prep}), we investigate the orbital characteristics of the interacting pairs sample. This work will provide a realistic set of parameters from which to produce the initial conditions of future idealized simulations, including the eccentricities, inclinations, and first pericentric separations. Additionally, we will assess orbital stability and the validity of the Keplerian approximation.

\section*{Acknowledgements}
KB acknowledges partial support from the National Science Foundation (NSF) Graduate Research Fellowship under Grant No. DGE-1329626. Support for JM is provided by the NSF (AST Award Number 1516374), by the Australian Research Council (ARC), and by the Harvard Institute for Theory and Computation, through their Visiting Scholars Program. FM acknowledges support through the Program ``Rita Levi Montalcini'' of the Italian MIUR. MV acknowledges support through an MIT RSC award, a Kavli Research Investment Fund, NASA ATP grant NNX17AG29G, and NSF grants AST-1814053 and AST-1814259. The authors would like to thank the anonymous reviewer, and the Scientific Editor Prof. Joop Schaye, both of whom provided invaluable feedback on the manuscript. In addition, we thank Dylan Nelson and Annalisa Pillepich for their thoughtful comments. The simulation used in this work, TNG100-1, is one of the flagship runs of the IllustrisTNG project, and was run on the HazelHen Cray XC40-system at the High Performance Computing Center Stuttgart as part of project GCS-ILLU of the Gauss Centre for Supercomputing.

\bibliographystyle{mnras}
\bibliography{GMTNG100_bib}{}

\begin{appendix}
\section{Technical Details}

\subsection{Data Structure}
\label{sec:App1}
The IllustrisTNG model is run at three different volumes (TNG50, TNG100 and TNG5300), each with a dark matter only run, and a dark plus baryonic matter run. Additionally, there are three (TNG100 and TNG300) or four (TNG50) iterations for each simulation that correspond to different initial conditions and resolutions. All simulations contain 100 nearly logarithmically spaced snapshots that span a redshift range of z = $[0 - 20]$. Particle data is available for all snapshots, and is organised based on three criteria: binding energy, subfind halo membership, and friend-of-friend (FoF) group membership. Subfind haloes (or, `subhaloes') are defined based on the {\sc subfind} algorithm \citep[][]{springel2001}, which links together baryonic and dark matter particles into locally over-dense and bound groups. The FoF haloes (or simply, `groups') are explicitly defined only for the dark matter particles using the FoF algorithm \citep{davis1985} with linking length $b=0.2$, however baryonic particles' membership to a FoF group is based on the membership of the closest dark matter particle.

\begin{table}
\resizebox{\columnwidth}{!}{%
\begin{tabular}{l|l?ll}
\multicolumn{2}{c?}{Numerical Parameters} & \multicolumn{2}{c}{Cosmological Parameters}  \\ \specialrule{.2em}{.1em}{.1em} 
Volume	& 110.7$^{3}$ Mpc$^{3}$	& \multicolumn{1}{l|}{$\Omega_{\rm dm}$} & 0.3089 \\
N$_{\rm gas}$	& 1820$^{3}$	& \multicolumn{1}{l|}{$\Omega_{\rm baryon}$} & 0.0486 \\
N$_{\rm dm}$	& 1020$^{3}$	& \multicolumn{1}{l|}{$\Omega_{\Lambda}$}	& 0.6911 \\
m$_{\rm baryon}$	& $1.4 \times 10^{6}$ M$_{\odot}$ & \multicolumn{1}{l|}{$h$}                   & 0.6744 \\
m$_{\rm dm}$	& $7.6 \times 10^{6}$ M$_{\odot}$ &	&	\\
$\epsilon_{\rm baryon}$ & 0.7 kpc	&	&	\\
$\epsilon_{\rm dm}$	& 1.4 kpc	&	&       
\end{tabular}%
}
\caption{\textit{Left:} Numerical specifications of TNG100-1. \textit{Right:} Cosmological parameters used in the IllustrisTNG model.}
\end{table}

\subsection{Merger Trees}
Merger trees \citep[][]{rod2015} use baryonic information within subfind haloes to trace mergers as a function of time. Such a merger tree is constructed by using three fundamental links: the Descendant, First Progenitor (FP), and Next Progenitor (NP). For a schematic of this network, refer to Figure 4 from \cite{nelson2015}. The descendant link tracks subfind haloes through time. The FP is the subfind halo with the largest mass history (i.e. the sum of this subhalo's progenitor masses along the main branch) of a given Descendant. The NP has the next largest mass history subfind halo of a that Descendant. A merger occurs when two subfind haloes share a Descendant. Put another way, a merger occurs when a Descendant has both a First and Next Progenitor. When parsing a merger tree, it is often useful to consider only the Main Progenitor Branch (MPB), which can be considered the ``trunk'' of the tree. This provides information only directly linked to the MPB. Parsing a merger tree requires at least two identifiers: the identification (ID) numbers of the First and Next Progenitors. Walking back along the MPB, each First Progenitor (FP) is defined by its index in the subhalo catalog at that snapshot until a FP can no longer be defined. For each FP, there is a network of Next Progenitors (NP) which were involved in a merger. Similarly, we terminate the merger tree traversal when there are no more Next Progenitors for a given FP.

\section{Visual Examples}
\label{sec:appb}
In this Appendix, we present our ideal mock SDSS $g$-band images for each of the 85 interacting pairs. These are organized roughly by their FoF group mass, with the most massive haloes at the top of the figure. 

\begin{figure*}
	\includegraphics[width=\textwidth]{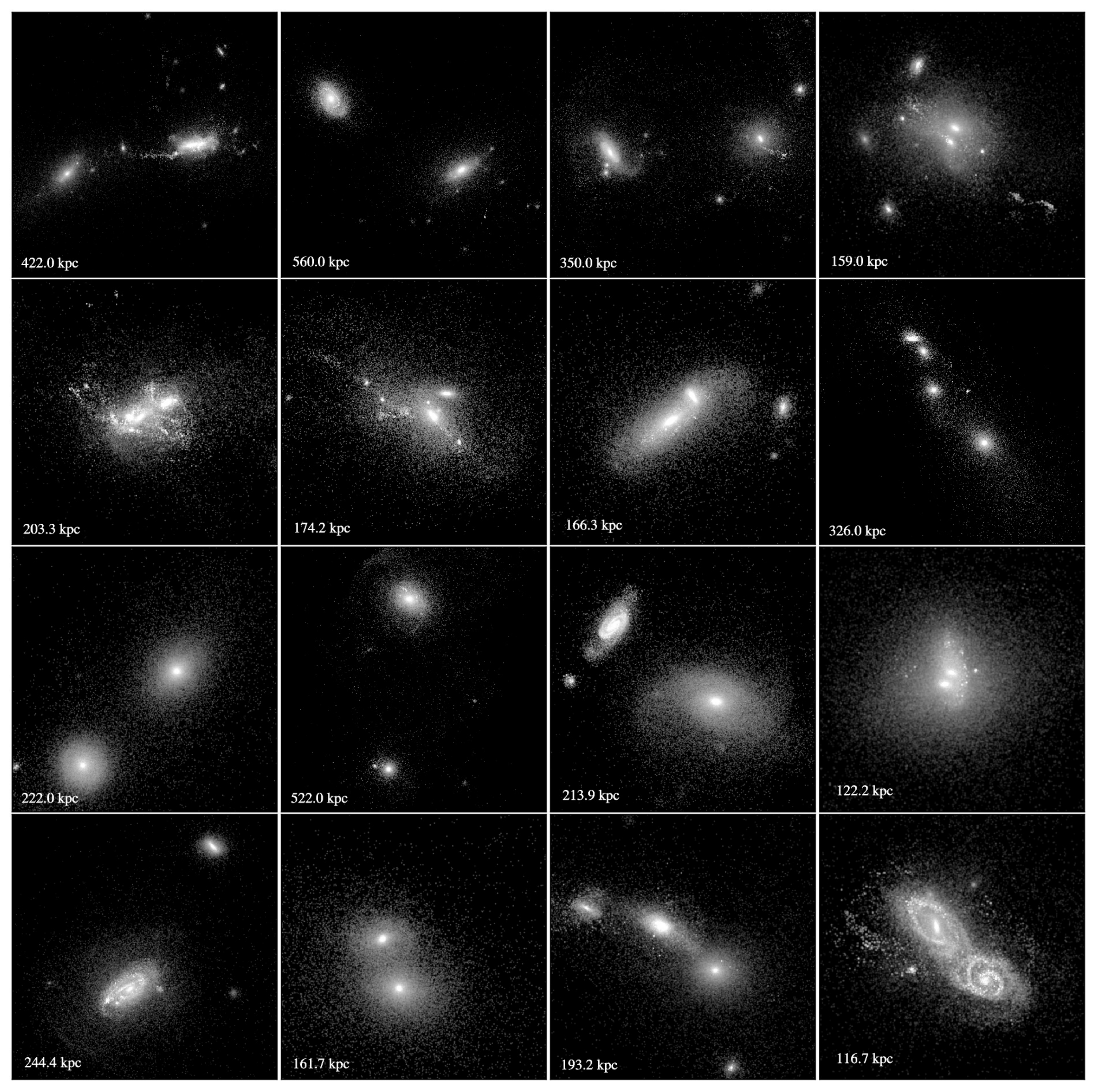}
\end{figure*}
\begin{figure*}
	\includegraphics[width=\textwidth]{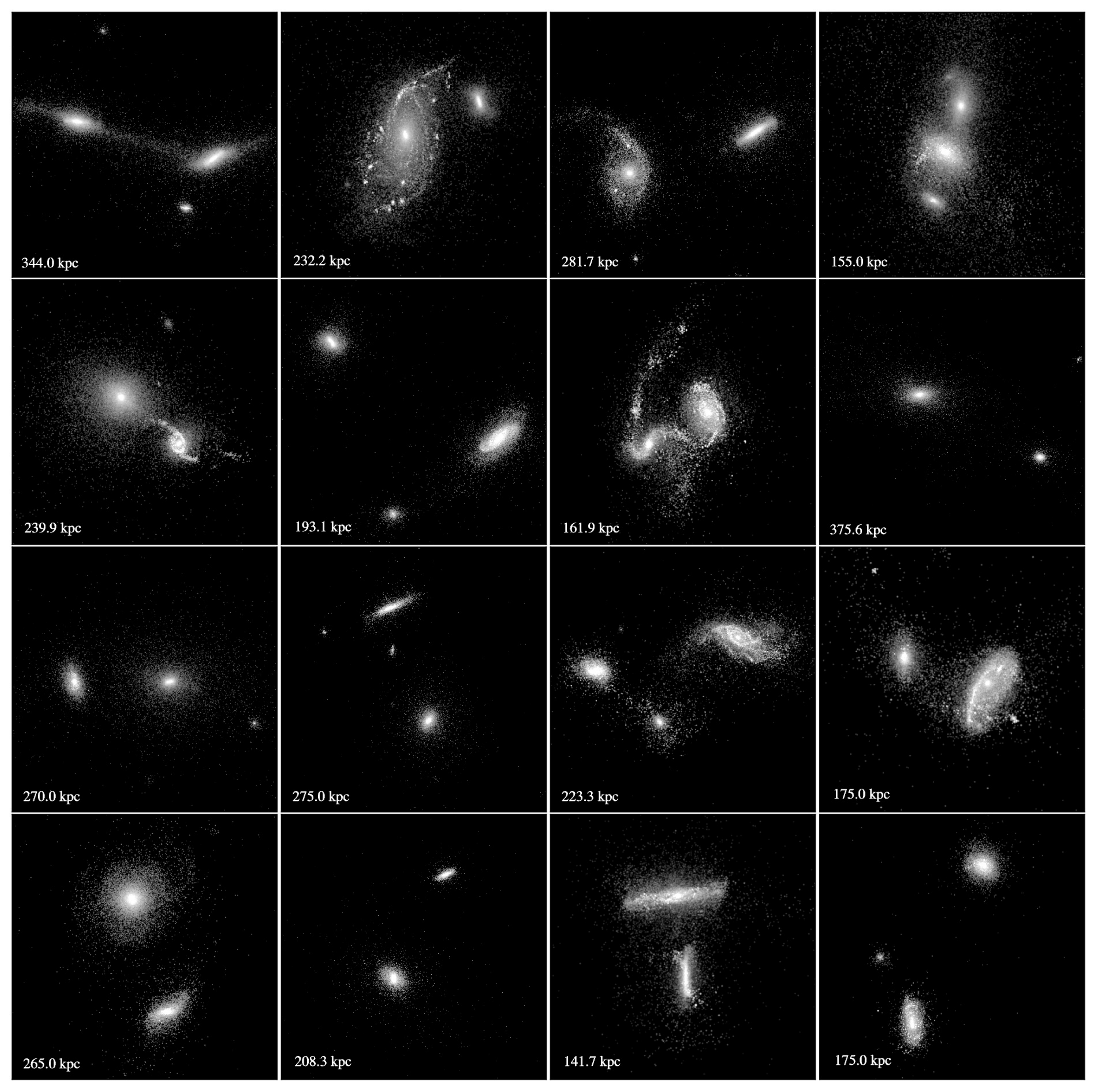}
\end{figure*}
\begin{figure*}
	\includegraphics[width=\textwidth]{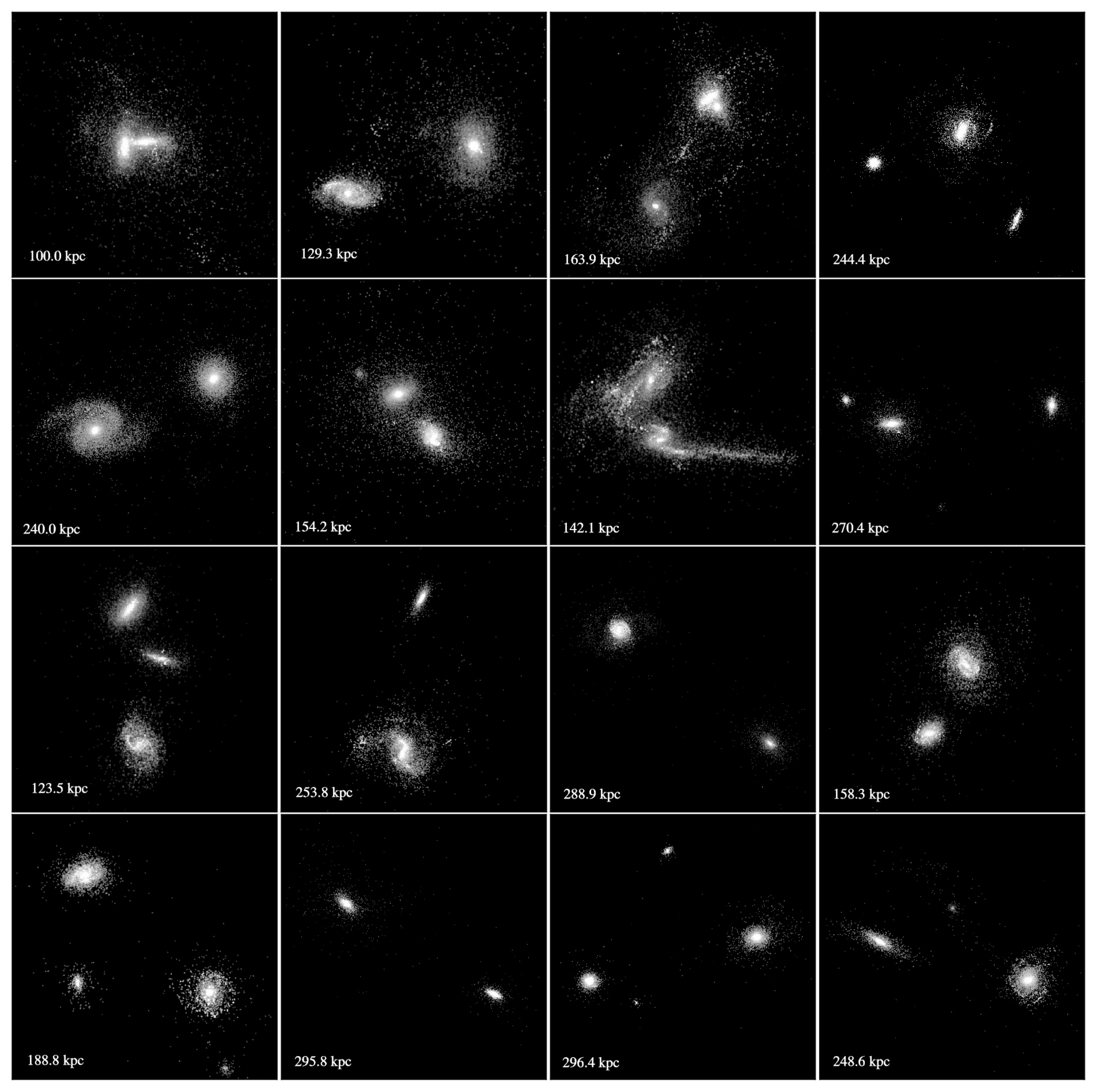}
\end{figure*}
\begin{figure*}
	\includegraphics[width=\textwidth]{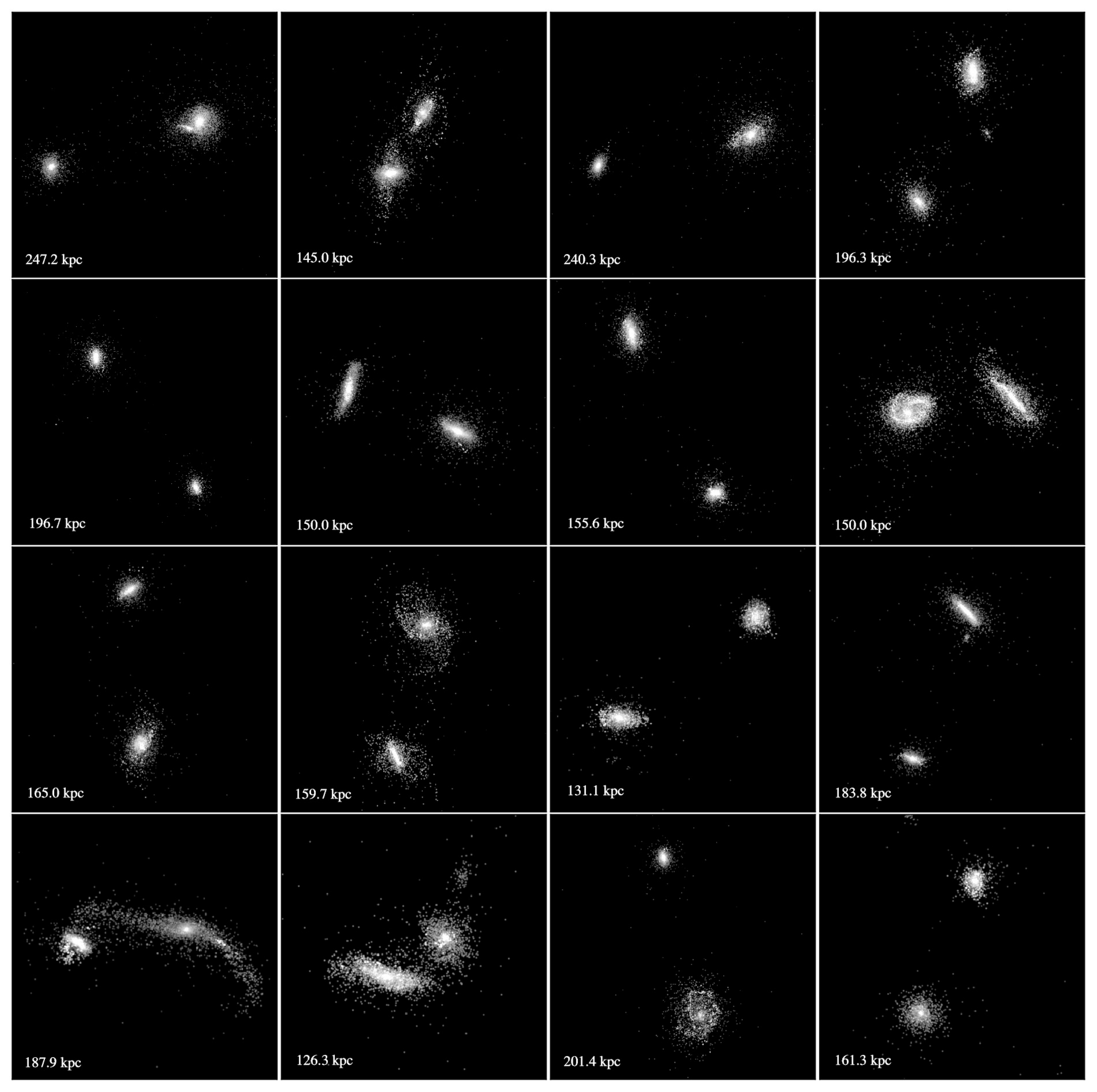}
\end{figure*}
\begin{figure*}
	\includegraphics[width=\textwidth]{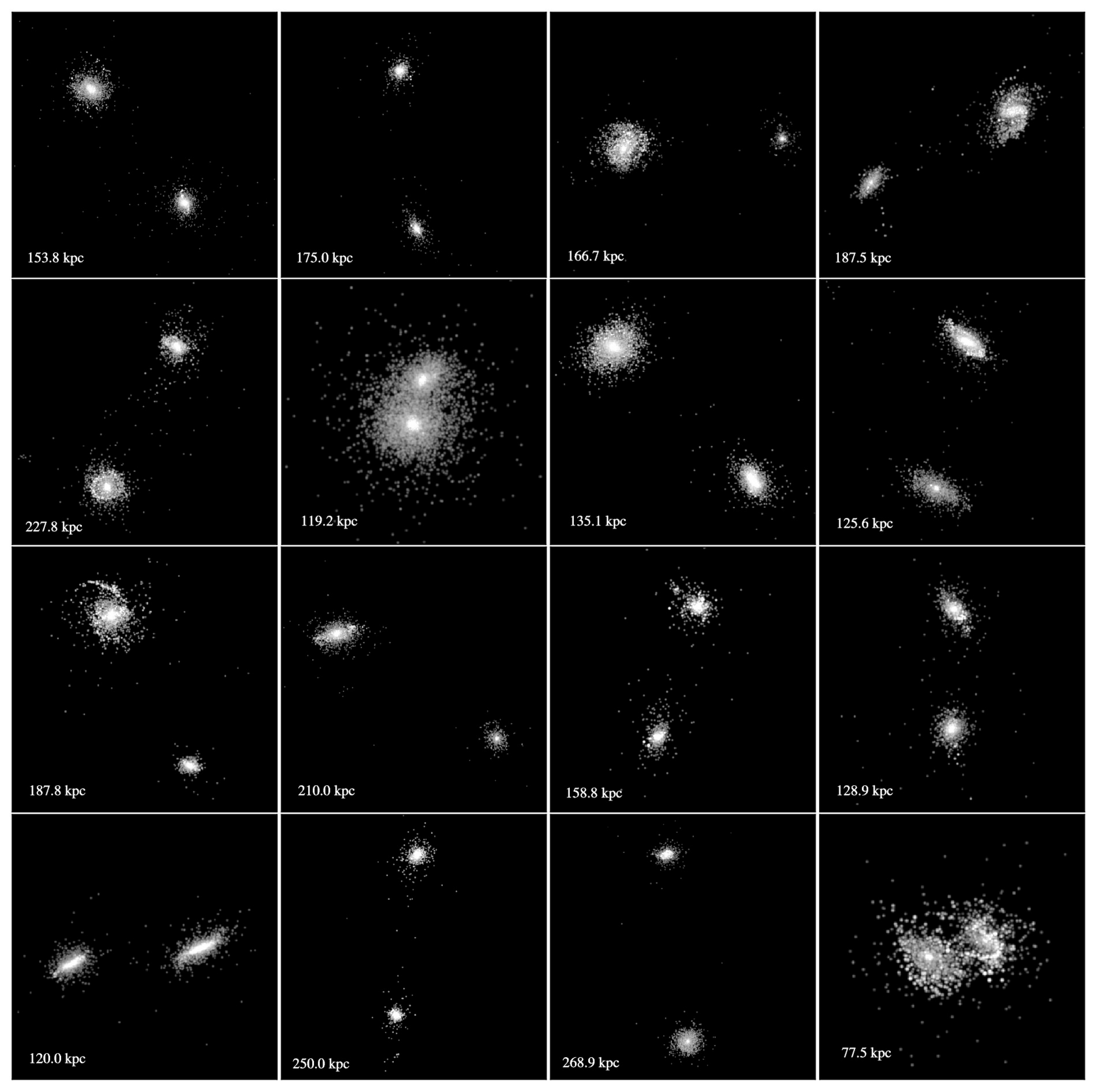}
\end{figure*}
\begin{figure*}
	\includegraphics[width=\textwidth]{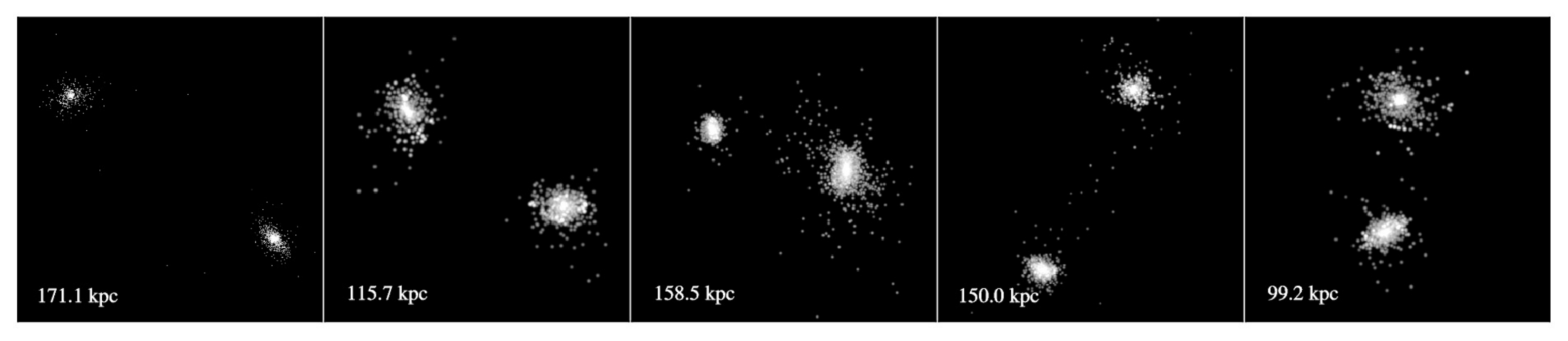}
	\caption{Ideal mock SDSS images of the TNG100-1 interacting pairs at $z=0$. Galaxies are ordered roughly by their FoF group mass, with the most massive haloes at the beginning, and the less massive haloes toward the end.}
	\label{fig:grid}
\end{figure*}
\end{appendix}

\bsp	
\label{lastpage}

\end{document}